\documentclass{elsart}
\usepackage[applemac]{inputenc}
\usepackage{graphicx}
\usepackage{amsmath}
\usepackage{times}
\usepackage{txfonts}
\usepackage{array}
\usepackage{natbib}
\usepackage{rotating}
\usepackage[dvipsnames,usenames]{color}
\usepackage{setspace}
\usepackage{setspace}

\newcommand{\te}[1]{10^{#1}}
\newcommand{\ut}[1]{\hspace{1mm}\mathrm{#1}}
\newcommand{\vc}[1]{\boldsymbol{\mathrm #1}}
\newcommand{\vecu}{\vc{u}} 
\newcommand{\vecB}{\vc{B}}

\newcommand{\cit}[1]{\let\temp=\\ \centering #1\let\\=\temp}
\newcommand{\rev}[1]{{#1}}

\newcolumntype{L}[1]{>{\raggedright\let\newline\\\arraybackslash\hspace{0pt}}m{#1}}
\newcolumntype{C}[1]{>{\centering\let\newline\\\arraybackslash\hspace{0pt}}m{#1}}
\newcolumntype{R}[1]{>{\raggedleft\let\newline\\\arraybackslash\hspace{0pt}}m{#1}}

\begin{document}

\begin{frontmatter}
\title{The interplay of fast waves and slow convection in geodynamo simulations nearing Earth's core conditions}
\author{Julien Aubert$^{1}$ and Nicolas Gillet$^{2}$}
\address{$^{1}$ Université de Paris, Institut de physique du globe de Paris, CNRS, F-75005 Paris, France.\\
$^{2}$ CNRS, ISTerre, University of Grenoble Alpes, Grenoble, France}

\begin{abstract}
Ground observatory and satellite-based determinations of temporal variations in the geomagnetic field probe a decadal to annual time scale range where Earth's core slow, inertialess convective motions and rapidly propagating, inertia-bearing hydromagnetic waves are in interplay. Here we numerically model and  jointly investigate these two important features with the help of a geodynamo simulation that (to date) is the closest to the dynamical regime of Earth's core. This model also considerably enlarges the scope of a previous asymptotic scaling analysis, which in turn strengthens the relevance of the approach to describe Earth's core dynamics. Three classes of hydrodynamic and hydromagnetic waves are identified in the model output, all with propagation velocity largely exceeding that of convective advection: axisymmetric, geostrophic Alfvén torsional waves, and non-axisymmetric, quasi-geostrophic Alfvén and Rossby waves. The contribution of these waves to the geomagnetic acceleration amounts to an enrichment and flattening of its energy density spectral profile at decadal time scales, thereby providing a constraint on the extent of the $f^{-4}$ range observed in the geomagnetic frequency power spectrum. As the model approaches Earth's core conditions, this spectral broadening arises because \rev{the decreasing inertia allows for waves at increasing frequencies. Through non-linear energy transfers with convection underlain by Lorentz stresses, these waves also} extract an increasing amount of energy from the underlying convection as their key time scale decreases towards a realistic value. The flow and magnetic acceleration energies carried by waves both linearly increase with the ratio of the magnetic diffusion time scale to the Alfvén time scale, highlighting the dominance of Alfvén waves in the signal and the stabilising control of magnetic dissipation at non-axisymmetric scales. Extrapolation of the results to Earth's core conditions supports the detectability of Alfvén waves in geomagnetic observations, either as axisymmetric torsional oscillations or through the geomagnetic jerks caused by non-axisymmetric waves. In contrast, Rossby waves appear to be too fast and carry too little magnetic energy to be detectable in geomagnetic acceleration signals of limited spatio-temporal resolution.
\end{abstract}


\end{frontmatter}

\section{\label{intro}Introduction}
The geomagnetic signal emanating from Earth's convecting fluid outer core contains energy over a wide range of time scales, from the longest paleomagnetic features on hundreds of million years down to the shortest variations observed over years and less at observatories and in satellite surveys. Throughout this range, time series of the palaeomagnetic dipole \citep{Constable2005,Panovska2013} as well as series of the field components at ground observatories \citep{DeSantis2003,Lesur2018} support the existence of several power-law ranges in the geomagnetic frequency power spectrum. At centennial periods and longer, synthetic time series produced by numerical simulations of the geodynamo have been used to relate the properties of this spectrum to key features of Earth's core geodynamics and magnetohydrodynamics \citep{Olson2012} and to calibrate stochastic evolution models describing these processes \citep{Buffett2015,Meduri2016}. At shorter time scales, within the range of geomagnetic secular variation that spans periods from centuries down to years, it has however been challenging to follow the same kind of approach because of the traditional limitations of geodynamo simulations. 

The situation has improved recently with a new generation of models \citep[see a recent review in][]{Wicht2019} working in a regime closer to the physical conditions of the core. A key physical time scale in the secular variation range is the core overturn time $\tau_{U}=D/U\approx 130 \ut{years}$, estimated using the core thickness $D=2260\ut{km}$ and a characteristic core surface fluid velocity $U=17 \ut{km/yr}$. Comparison with numerical simulations \citep{Bouligand2016,Aubert2018} has revealed that this time scale relates to a corner frequency in the geomagnetic field power spectrum, and marks the start of a high-frequency range with a spectrum following the power law $P\sim f^{-4}$ (with $f$ the frequency) similar to that inferred from observatory time series \citep{DeSantis2003}. This range may extend up to frequencies of about $1\ut{yr^{-1}}$ \citep{Lesur2018}, though the separation of the internally generated geomagnetic field from external sources is already difficult at this point. At such high frequencies, fast hydromagnetic waves are expected to be present in the geomagnetic signal. Recent advances in geomagnetic field modelling have for instance enabled the retrieval of torsional Alfvén waves in the core \citep{Gillet2010,Gillet2015}, at a fundamental period close to 6 years. Other types of Alfvén waves also appear to play a key role in the description of geomagnetic jerks \citep{Aubert2019}. The main goal of this study is to explore the role of these waves in structuring the geomagnetic power spectrum at high frequencies. More generally, we wish to understand the interplay between fast hydromagnetic waves and the slower convection of Earth's core. This is also essential for linking the rapid geomagnetic signals now routinely observed with satellites to the physical properties of Earth's core, and for the prospect of developing better geomagnetic predictions at horizons of human interest.

This problem may be viewed from a different angle by associating the time scales relevant to the geomagnetic variations to a hierarchy of force balances in the core. It has long remained difficult to grasp the place taken by the Lorentz force in this hierarchy, because of the non-linear nature of this force and the self-sustained character of the magnetic field. The situation has however been largely clarified recently, following theoretical advances \citep{Davidson2013,Calkins2015,Aurnou2017,Calkins2018} and advancing numerical explorations of the geodynamo simulation parameter space \citep{Aubert2017,Schaeffer2017,Aubert2019b,Schwaiger2019,Schwaiger2021}. Throughout this parameter space indeed, the rapid planetary rotation implies that at the system scale, the leading-order equilibrium in amplitude occurs between the Coriolis and pressure forces. In most models, this quasi-geostrophic (QG) equilibrium is perturbed at the next order with a triple MAC balance between the buoyancy, Lorentz and residual ageostrophic component of the Coriolis force. In conditions approaching those of Earth's core, the MAC balance is exactly satisfied at a rather large scale $d_{\perp} \sim D/10$, where magnetic energy is fed into the system from the buoyancy force \citep{Aubert2019b}. Writing the QG-MAC balance yields two other independent estimates of the core overturn time $\tau_{U}$ \citep[e.g.][]{Davidson2013}
\begin{equation}
\tau_{U} = \dfrac{D}{U} \approx \dfrac{\rho \Omega d_{\perp}}{g_{o} C} \approx  \dfrac{\rho \mu \Omega d_{\perp}^{2}}{B^2}.\label{MACts}
\end{equation}
Using typical values for the rotation rate $\Omega=7.29~\te{-5}\ut{s^{-1}}$, density $\rho=11000\ut{kg/m^3}$, magnetic permeability $\mu=4\pi\te{-7}\ut{H/m}$, core surface gravity $g_{o}=10\ut{m/s^2}$, convective density anomaly $C=\te{-5}\ut{kg/m^3}$ \citep[e.g.][]{Jones2015,Aubert2020} and magnetic field strength $B=4\ut{mT}$ \citep{Gillet2010}, both estimates lead to $\tau_{U}=\mathcal{O}(\te{2})\ut{yr}$, consistent with the first direct estimate mentioned above. Through the QG-MAC balance, the core overturn time scale is therefore that of the slow convective evolution of the velocity field, adjusting in an inertia-less manner to the creation and advection of density anomaly and magnetic field structures. A range of length scales and amplitudes is obviously expected for these structures, such that a broader range of convective time scales is expected around $\tau_{U}$, which appears open-ended on the long side but bound to decadal periods on the short side \citep{Aubert2018}. 

While recent advanced dynamo models largely preserve the dynamics described by earlier models on time scales of centuries and longer \citep{Aubert2017,Aubert2018}, their strength and purpose is to render the interannual and decadal range of time scales much shorter than $\tau_{U}$, where the dynamics can become inertial again and take the form of rapid responses to transient disruptions in the QG-MAC balance. At conditions approaching those of Earth's core, inertia indeed comes several orders of magnitude below the MAC forces \citep{Aubert2019b}, such that the flows and magnetic signals associated with these disruptions are expected to remain small respectively to the convective flow and main magnetic field. This should naturally lead to quasi-linear perturbations in the governing equations and, in the presence of restoring forces, wave propagation. \rev{The rapid, inertia-bearing (I) waves that have been theoretically investigated \citep[e.g.][]{Finlay2008houches} also mainly involve the magnetic (M) and Coriolis (C) forces (with buoyancy sustaining rapid waves only in contexts of stable stratification). At the axisymmetric level, the Coriolis force identically vanishes on axial cylindrical surfaces, leading to purely MI torsional waves that represent} a purely geostrophic case of Alfvén waves. These have been identified in Earth's core by \cite{Gillet2010,Gillet2015} and have become increasingly clear in numerical simulations as the model parameters have become more realistic \citep{Wicht2010,Teed2014,Schaeffer2017,Aubert2018}. The relevant time scale for torsional waves is the Alfvén time
\begin{equation}
\tau_{A}=\dfrac{D\sqrt{\rho\mu}}{B},
\end{equation}
which, using the estimates quoted above, indeed satisfies $\tau_{A}\approx 2\ut{yr}\ll\tau_{U}$ in Earth's core. \rev{At the non-axisymmetric level, the Coriolis force in principle takes over the inertial force \citep{Lehnert1954,Hide1966,Finlay2010,Labbe2015}, leading to slow MC modes that do not contribute to rapid dynamics. In interesting recent developments, it has however been shown that the influence of the Coriolis force can be mitigated in a variety of ways, leading to fast, quasi-geostrophic (axially columnar), non-axisymmetric dynamics. Starting their survey from low-frequency MC modes, \cite{Gerick2020} have identified modes approaching the Alfvén frequency $1/\tau_{A}$ from below as their radial complexity is increased. Their kinetic to magnetic energy ratio also increases, meaning that the influence of inertia is gradually restored. In advanced numerical dynamo simulations that may be seen as an extreme case of spatial complexity, \cite{Aubert2018} have highlighted a spatial segregation of the force balance. In regions where the magnetic field is relatively homogeneous, localised and nearly MI Alfvén wave dynamics is observed with frequencies also closely approaching the Alfvén frequency from below. The MC part of the balance remains confined close to regions of strong magnetic field heterogeneity. Despite the constraints set by the Coriolis force, MI Alfvén waves therefore appear to be possible at any spatial scale, but detecting these} in the geomagnetic signal is a challenge as \rev{they evolve close to, or at kinetic to magnetic energy equipartition.} Their magnetic signature is therefore nominally small respectively to the main magnetic field produced by convection, the energy of which exceeds the total kinetic energy in Earth's core by a factor $A^{-2}=(\tau_{U}/\tau_{A})^{2} \approx 5000$. In this respect, wave focusing below the core-mantle boundary has recently been proposed as a way to overcome this challenge and also to explain the signature of recent geomagnetic jerks \citep{Aubert2019}. 

\rev{At frequencies in excess of the Alfvén frequency and up to twice the rotation rate $\Omega$ (on the order of an Earth day), CI inertial waves populate}  a discrete and dense frequency spectrum \citep{Rieutord1997}. The issue of detectability may be seen as even worse here \rev{as the kinetic to magnetic energy ratio of inertial waves further increases with increasing frequency \citep{Gerick2020}, such that they weakly interact with the magnetic field.} There remains however the possibility of slow, quasi-geostrophic 'Rossby' modes \citep{Zhang2001,Busse2005,Canet2014} with a pulsation $\omega\ll \Omega$ approaching the Alfvén frequency from above as the spatial complexity is increased, which enter the realm of interannual geomagnetic signals and also \rev{carry higher amounts of magnetic energy}. The second goal of this study is to use our models to better quantify the detectability of Alfvén and slow Rossby waves in Earth's core.

We report here on a new numerical geodynamo model that enables a joint exploration of all these waves and convective features at an unprecedentedly realistic level of separation between the key time scales $\tau_{U}$, $\tau_{A}$ and $\tau_\Omega$. Being to date the closest to Earth's core conditions, this model also considerably enlarges the parameter range over which an asymptotic scaling analysis relevant to Earth's core can be performed, which forms the third goal of this study. The manuscript is organised as follows: section \ref{model} presents the existing path theory and the new numerical model. Results are presented in section \ref{results} and discussed in section \ref{discu}. 

\section{\label{model}Model and Methods}
\subsection{Model set-up and numerical method}
We use a standard numerical model for Boussinesq convection, thermochemical density anomaly transport, and magnetic induction in the magnetohydrodynamic approximation. Full details on the equation set, boundary conditions and numerical method can be found in \cite{Aubert2013b,Aubert2017} and \cite{Aubert2018}, where the specific configuration is denoted as 'CE' (Coupled Earth). The unknown fields are the velocity field $\vecu$, magnetic field $\vecB$ and density anomaly field $C$. The fluid domain is an electrically conducting and rotating spherical shell of thickness $D=r_{o}-r_{i}$ and aspect ratio $r_{i}/r_{o}=0.35$ representing the Earth's outer core. The shell takes place between an inner core of radius $r_{i}$, and a solid mantle between radii $r_{o}$ and $r_\mathrm{E}=1.83 r_{o}$ (the surface of the Earth), both of which are conducting and electromagnetically coupled to the fluid shell. The electrical conductivity of the inner core is set to the same value $\sigma$ as that of the outer core, while the mantle features a thin conducting layer at its base of thickness $\Delta$ and conductivity $\sigma_{m}$ such that the ratio of conductances is $\Delta\sigma_{m}/D\sigma=\te{-4}$. The inner core is furthermore  gravitationally coupled to the mantle, and the three layers can present axial differential rotations with respect to each other, with the constant axial rotation of the ensemble defining the planetary rotation vector $\vc{\Omega}=\Omega\vc{e}_{z}$. Moments of inertia for these three layers respect their Earth-like proportions. A homogeneous mass anomaly flux $F$ imposed at radius $r=r_{i}$ drives convection from below, and the mass anomaly flux vanishes at $r=r_{o}$ (neutral buoyancy beneath the outer surface). A volumetric buoyancy sink is present in the volume to ensure mass conservation. On top of this homogeneous buoyancy distribution, lateral heterogeneities in the mass anomaly flux are superimposed at $r=r_{i}$ and $r=r_{o}$, respecting the 'CE' setup. Stress-free mechanical boundary conditions are also imposed on the fluid shell. This removes the need to resolve extremely thin viscous boundary layer that have a negligible influence on the solution \citep{Aubert2017}. As in our previous work, all models presented here produced a self-sustained, dipole dominated magnetic field of Earth-like geometry \citep{Christensen2010} that did not reverse polarity during the integration time. 

We consider a spherical coordinate system $(r,\theta,\varphi)$ with unit vectors $\vc{e}_{r},\vc{e}_{\theta},\vc{e}_{\varphi}$ associated to the cylindrical system $(s,z,\varphi)$ of unit vectors $\vc{e}_{s},\vc{e}_{z},\vc{e}_{\varphi}$. The numerical implementation involves a decomposition of $\vecu$, $\vecB$ and $C$ in spherical harmonics up to degree and order $\ell_{\mathrm{max}}$ and a discretisation in the radial direction on a second-order finite-differencing scheme with $NR$ grid points. We use the spherical harmonics transform library  SHTns \citep{Schaeffer2013} freely available at {\tt https://bitbucket.org} {\tt/nschaeff}{\tt/shtns}. We use a second-order, semi-implicit time stepping scheme. The solution is approximated using smoothly ramping hyperdiffusivity applied on the velocity and density anomaly fields, but not on the magnetic field that remains fully resolved. The functional form of hyperdiffusivity involves a cut-off $\ell_{h}=30$ below which hyperdiffusivity is not applied, and an exponential ramping parameter $q_{h}$ such that $(\nu_\text{eff},\kappa_\text{eff})=(\nu,\kappa)\, q_{h}^{\ell-\ell_{h}}$, describing the increase of hyperdiffusivity with spherical harmonic degree $\ell$ for $\ell\ge\ell_{h}$. When compared against fully resolved references \citep{Aubert2019b}, the approximated solutions adequately preserve the QG-MAC balance and the large-scale morphology of the solution. They enable the computation of long temporal sequences that would otherwise not be feasible and are therefore tailored towards the present analysis of dynamics in the time domain. Values of $NR$, $\ell_\mathrm{max}$ and $q_{h}$ used in our models are reported in Table \ref{inputs}.

\subsection{Path theory, dimensionless inputs and outputs}
We recall here the four main parameters of the model, the flux-based Rayleigh, Ekman, Prandtl and magnetic Prandtl numbers:
\begin{equation}
Ra_{F}=\dfrac{g_{o}F}{4\pi\rho\Omega^{3}D^{4}},~ E=\dfrac{\nu}{\Omega D^{2}},~ Pr=\dfrac{\nu}{\kappa},~ Pm=\dfrac{\nu}{\eta}.
\end{equation}
Aside of the already introduced planetary rotation rate $\Omega$, core thickness $D$, core surface gravity field $g_{o}$, core density $\rho$, bottom-driven mass anomaly flux $F$, these expressions involve the viscous, thermo-chemical and magnetic diffusivities $\nu$, $\kappa$ and $\eta$ (with $\eta=1/\mu\sigma$). \rev{Details on the correspondance between flux-based and canonical Rayleigh numbers may be found in \cite{ChristensenAubert2006}}. Our main model case in this study uses the path theory \citep{Aubert2017} that bridges the parameter space gap between our previous coupled Earth model \citep{Aubert2013b} and Earth's core conditions by relating these four parameters to a single variable $\epsilon$:
\begin{eqnarray}
Ra_{F}(\epsilon)=\epsilon Ra_{F} (1),~ E(\epsilon)=\epsilon E (1),~ Pr(\epsilon)=1,~ Pm(\epsilon)=\sqrt{\epsilon} Pm (1).
\end{eqnarray}
Here $Ra_{F}(\mathrm{1})=2.7~\te{-5}$, $E(\mathrm{1})=3~\te{-5}$ and $Pm(\mathrm{1})=2.5$ are the control parameters of the coupled Earth dynamo model. \rev{The choice of a constant $Pr=1$ along the path stems from the thermal and chemical density anomalies being blended together in a co-density approach, and is also relevant given that hyperdiffusivity of equal strength is applied to the velocity and density anomaly fields.} Previously \citep{Aubert2017,Aubert2018}, we have explored this path down to $\epsilon=3.33~\te{-4}$, halfway in a logarithmic progression scale between the start ($\epsilon=1$) and the Earth's core conditions located at $\epsilon=\te{-7}$. Here we mainly report on a model at $\epsilon=\te{-5}$, at 71 percent of this path on the same logarithmic scale (from hereafter the 71p model). Table \ref{inputs} reports on the input parameters of this main model case, and of the other cases used in this study. To date, the 71p model is the closest to Earth's core conditions that has been reached in a numerical geodynamo simulation, in particular regarding the nearly inviscid behaviour achieved at large scales, as witnessed by the Ekman number $E=3~\te{-10}$. \rev{This is of course made possible by the hyperdiffusive approximation and the use of stress-free boundaries. For comparison, the fully resolved numerical simulation currently closest to core conditions is that of \cite{Schaeffer2017}, operating at $E=3~\te{-7}$.}
 
\begin{table}
\caption{\label{inputs} Input parameters of numerical models (see text for definitions). All models have $Pr=1$. Model cases with a star have been previously published in \cite{Aubert2017,Aubert2018,Aubert2019}}
\vspace*{\baselineskip}

\begin{center}
\setstretch{0.8}\small
\hspace*{-1cm}\begin{tabular}{lrm{1.5cm}m{1.2cm}rrrrm{0.5cm}m{0.5cm}}
\hline\\[-0.5cm]
Case & $\epsilon$ & Path position (percent) & run length ($\tau_{U}$ units) & $E$ & $Ra_{F}$ & $Pm$ & $NR$ & $\ell_\mathrm{max}$ & $q_{h}$\\
\hline
29p* & $\te{-2}$ & 29 & 332 & $3~\te{-7}$ & $2.7 ~\te{-7}$ & 0.25 & 320 & 133 & 1.07\\
Midpath* & $3.33~\te{-4}$ & 50 & 196 & $\te{-8}$ & $9 ~\te{-9}$ & 0.045 & 624 & 133 & 1.10\\
71p & $\te{-5}$ & 71  & 117 & $3~\te{-10}$ & $2.7~\te{-10}$ & $7.9~\te{-3}$ & 1248 & 133 170 & 1.14 1.09 \\
50pRa+ & & off-path & 109 & $\te{-8}$ & $3.6~\te{-8}$ & 0.045 & 624 & 133  & 1.13 \\
50pRa+Pm- & & off-path & 68 & $\te{-8}$ & $3.6~\te{-8}$ & 0.023 & 624 & 133  & 1.13 \\
50pPm+ & & off-path & 37 & $\te{-8}$ & $9~\te{-9}$ & 0.1125 & 624 & 133  & 1.10\\
\hline
\end{tabular}
\end{center}
\end{table}

In our previous analyses that stopped at 50 percent of the path \citep[the so-called Midpath model,][]{Aubert2017,Aubert2018,Aubert2019}, it was found that many of the important diagnostics related to waves evolve rather subtly, and scale with weak powers of $\epsilon$. To reach our main goals, it is therefore necessary to cover a wide range of this parameter. This is the main reason that motivated a direct leap from 50 to 71 percent of the path here, rather than the progression in half-decades of $\epsilon$ that we favoured previously. The 71p model has been initialised by taking a state from the Midpath model. Initial transients could almost be eliminated by applying a new approach. First we applied the path scaling laws \citep{Aubert2017} to cast the amplitudes of $\vecu,\vecB,$ and $C$ towards predictions close to their actual values at 71 percent of the path. The numerical mesh was then refined to achieve $NR=1248$ and $\ell_{max}=133$. Finally the computation was restarted with the physical parameters relevant to 71 percent of the path and hyperdiffusivity parameters $\ell_{h}=30$ and $q_{h}=1.14$. After 9.5 overturn times $\tau_{U}$, the lateral resolution has been increased to $\ell_{max}=170$, enabling a partial relaxation of the hyperdiffusivity strength down to $q_{h}=1.09$ for the rest of the computation (about 107.5 overturns). Three secondary models have also been derived from the Midpath model by changing its Rayleigh number (label Ra+) or its magnetic Prandtl number (labels Pm+/-). These models complement the series computed along the path with a limited exploration across the path around the Midpath position. 

\begin{table}
\caption{\label{outputs1} Output parameters, presented in a. as dimensionless ratios of time scales, and in b. as dimensional time scales obtained by setting a dimensional value to the magnetic diffusion time $\tau_{\eta}$ (see text). Estimates for Earth's core are also reported in panel b \citep[taken from][]{Gillet2010,Lhuillier2011b,Christensen2012,Aubert2017,Aubert2018,Aubert2019}.}
\begin{center}
a.\vspace*{\baselineskip}

\hspace*{-1cm}\begin{tabular}{lrrrrrrrrrr}
\hline\\[-0.5cm]
Case & $Ro=\dfrac{\tau_{\Omega}}{\tau_{U}}$ & $\lambda=\dfrac{\tau_{\Omega}}{\tau_{A}}$ & $Rm=\dfrac{\tau_\eta}{\tau_{U}}$ & $A=\dfrac{\tau_{A}}{\tau_{U}}$ & $S=\dfrac{\tau_{\eta}}{\tau_{A}}$ & $\dfrac{\tau_{\mathrm{SV}}^{1}}{\tau_{\eta}}$ & $\dfrac{\tau_{\mathrm{SA}}^{0}}{\tau_{\eta}}$\\[0.3cm]
\hline
29p* & $1.26~\te{-3}$ & $5.15~\te{-3}$ & 1046 & 0.244 & 4286 & $3.12~\te{-3}$ & $1.11~\te{-4}$ \\
Midpath* & $2.40~\te{-4}$ & $2.10~\te{-3}$ & 1082 & 0.114 & 9491 & $3.06~\te{-3}$ & $8.72~\te{-5}$\\
71p   & $4.31~\te{-5}$ & $8.82~\te{-4}$ & 1136 & 0.049 & 23234 & $2.89~\te{-3}$ & $7.88~\te{-5}$ \\
50pRa+  & $4.45~\te{-4}$ & $3.16~\te{-3}$ & 2003 & 0.141 & 14235 & $1.65~\te{-3}$ & $3.78~\te{-5}$ \\
50pRa+Pm- & $4.78~\te{-4}$ & $3.10~\te{-3}$ & 1101 & 0.155 & 7100 & $3.1~\te{-3}$ & $9.09~\te{-5}$ \\
50pPm+ & $2.22~\te{-4}$ & $2.37~\te{-3}$ & 2495 & 0.094 & 26671 & $1.32~\te{-3}$ & $2.62~\te{-5}$ \\
\hline
\end{tabular}
\vspace*{\baselineskip}

b.\vspace*{\baselineskip}

\hspace*{-1cm}\begin{tabular}{lrrrrrrrrrr}
\hline\\[-0.5cm]
Case & $\tau_{\eta}$ (yr) & $\tau_{U}$ (yr) & $\tau_{A}$ (yr) & $2\pi\tau_{\Omega}$ (days) & $\tau_{\mathrm{SV}}^{1}$ & $\tau_{\mathrm{SA}}^{0}$  \\[0.2cm]
\hline
29p* & 135000 & 129 & 31.5 & 373 & 421 & 15.0\\
Midpath* & 135000 & 125 & 14.3 & 69  & 413 & 11.8 \\
71p  & 135000 & 119 & 5.8 & 11.8 & 390 & 10.6\\
50pRa+  & 245600 & 123 & 17.3 & 125 & 405 & 9.3\\
50pRa+Pm- & 135000 & 123 & 19 & 134 & 419 & 12.3\\
50pPm+ & 305900 & 123 & 11.5 & 62 & 404 & 8.0\\
Earth & $6~\te{4}-3~\te{5}$ & $\approx 130$ & $\approx 2$ & 1 &  $\approx 415$ & $\approx 10$ \\
\hline
\end{tabular}
\end{center}
\end{table}

The root-mean-squared velocity and magnetic field amplitudes $U$ and $B$ measured in the shell over the course of the simulation give access to the overturn and Alfvén time scales $\tau_{U}=D/U$ and $\tau_{A}=D\sqrt{\rho\mu}/B$. These are presented in Table \ref{outputs1}a as dimensionless ratios taken respectively to the two other important time scales in the system, the rotational time scale $\tau_{\Omega}=1/\Omega$ and the magnetic diffusion time $\tau_{\eta}=D^{2}/\eta$. As we advance along the path towards Earth's core conditions (or decrease $\epsilon$), the numerical models preserve an approximately constant value of the magnetic Reynolds number $Rm=\tau_{\eta}/\tau_{U}\approx 1000$, while they enlarge the time scale separation between $\tau_{\eta},\tau_{U}$ with the Alfvén and rotational time scales. Time scale separation can be measured using the Alfvén number $A=\tau_{A}/\tau_{U}$, Lundquist number $S=\tau_{\eta}/\tau_{A}$, Rossby number $Ro=\tau_{\Omega}/\tau_{U}$ and Lehnert number $\lambda=\tau_{\Omega}/\tau_{A}$. Along the lines introduced by \cite{Jault2008}, \cite{Gillet2011} and \cite{Aubert2017}, the asymptotic regime relevant to the geodynamo may be formalised through the two conditions $\lambda \ll 1$ and $A \ll 1$, which in turn imply $Ro \ll 1$ and (if $Rm \gg 1$) $S \gg 1$. In \cite{Aubert2018} the entry into this regime has been located at 29 percent of the path, from the consideration of flow and magnetic acceleration signatures. The 71p model provides the largest and most realistic separation between the four important time scales $\tau_{\eta},\tau_{U},\tau_{A},\tau_{\Omega}$ that has been obtained to date in a geodynamo simulation. The unidimensional nature of the path however implies that all time scales are covarying, which may lead to difficulties in disentangling the exact effect of each of these scales. For this reason, the three other secondary models taken across the path close to Midpath conditions will be useful in the analysis presented in section \ref{results}. 

\subsection{\label{pathcomp}Additional dimensionless outputs for scaling along the path}
\begin{table}
\caption{\label{outputs2} Additional dimensionless outputs of the 71p model.}
\vspace*{\baselineskip}

\begin{center}
\begin{tabular}{rl}
\hline
Ohmic fraction $f_\mathrm{ohm}$ & 0.89 \\
magnetic diffusion length scale $d_\mathrm{min}/D$ & $1.82~\te{-2}$\\
Taylorisation level on axial cylinders $\mathcal{T}$ & $6.5~\te{-3}$\\
\hline
\end{tabular}
\end{center}
\end{table}

The computation of the 71p model also provides an opportunity to update the scaling analysis initially presented in \cite{Aubert2017}, which stopped at 50 percent of the path. For this purpose, here we list and report on the few additional time-averaged dimensionless outputs that are needed \citep[see][for full definitions and Table \ref{outputs2} for values]{Aubert2017}: the fraction $f_\mathrm{ohm}$ of convective power that is dissipated in Ohmic losses, the magnetic diffusion length scale $d_\mathrm{min}/D$, or square root of the ratio between time-averaged magnetic energy and Ohmic losses, and the time-averaged level of enforcement $\mathcal{T}$ of the Taylor constraint on axial cylindrical surfaces in the shell. As done at previous positions along the path \citep{Aubert2017,Aubert2019}, we also determine a scale-dependent representation of the force balance in the 71p model, where the root-mean-squared amplitude of each force is represented as a function of the spherical harmonic degree $\ell$. This scale-dependent force balance tool has previously revealed its strength in deciphering the complex hierarchy of forces present in the system, including the non-trivial roles of hydrodynamic and magnetic pressure, and in associating the successive balances coming along this hierarchy to physically grounded length scales \citep{Aubert2017,Aubert2019b,Schwaiger2019,Schwaiger2021}.

\subsection{\label{doutputs}Integral dimensional outputs}
For the purpose of geophysical applications, it has become common practice to cast the dimensionless outputs of the numerical model back into the dimensional world, in a way that is rationalised by the physical equilibria preserved along the path to Earth's core \citep[see e.g.][for a detailed discussion]{Aubert2020}. Length scales are dimensioned by setting $D=2260\ut{km}$ as Earth's outer core thickness. Along the path, an in agreement with our previously used conventions \citep{Aubert2019}, the time basis is provided by setting the magnetic diffusion time to $\tau_{\eta}=135000\ut{yr}$, corresponding to a magnetic diffusivity $\eta=1.2~\ut{m^2/s}$ at the mid-point of current estimates \citep{Aubert2017}. The resulting dimensional time scales values are summarised in Table \ref{outputs1}b. Together with the constant value $Rm\approx1000$ obtained along the path, this choice for $\tau_{\eta}$ ensures a geophysically realistic value $\tau_{U}\approx 130\ut{yr}$ for the convective overturn time. \rev{This also leads to} geophysically realistic values for two other important time scales pertaining to convectively driven geomagnetic signals, the secular variation and acceleration time scales $\tau_{\mathrm{SV}}^{1}\approx 400\ut{yr}$ and $\tau_{\mathrm{SA}}^{0}\approx 10\ut{yr}$ \citep[see][for definitions and discussion]{Lhuillier2011b, Christensen2012,Aubert2018}. This implies that all models taken along the path are adjusted for an adequate simulation of slow inertialess convection and convection-driven geomagnetic variations in the decadal to secular range. Turning now to the time scales pertaining to fast hydromagnetic waves, the progress made by the 71p model over previous efforts becomes tangible when we consider the dimensional value of the Alfvén time scale $\tau_{A}=5.8\ut{yr}$, now only a factor 3 away from the Earth value. Likewise, the duration of a physical day is $2\pi\tau_{\Omega}=11.8\ut{days}$ is now only an order of magnitude away from an Earth day. 

For the series of 50p models taken across the path, we also aim at preserving the realism of convection-driven geomagnetic signals i.e. also achieving realistic values for $\tau_{U}, \tau_{\mathrm{SV}}^{1}$ and $\tau_{\mathrm{SA}}^{0}$. This is straightforwardly achieved by adjusting $\tau_{\eta}$ in proportion of the magnetic Reynolds numbers achieved by these models. Model 50pRa+Pm- uses a combination of Rayleigh and magnetic Prandtl number such that $Rm$ keeps a value close to that of models along the path, so the same time basis $\tau_{\eta}=135000\ut{yr}$ is used. The time basis of models 50pRa+ and 50pPm+ is adjusted accordingly to the ratio between their magnetic Reynolds number and that of model 50pRa+Pm-, leading to the respective time bases $\tau_{\eta}=245600\ut{yr}$ and $\tau_{\eta}=305900\ut{yr}$. Interestingly, these two models also sample a geophysically relevant range for possible values of $\tau_{\eta}$ and $Rm$, particularly concerning the possibility of high thermal and electrical conductivity in the core \citep[e.g.][]{Pozzo2012}. 

As in our previous studies \citep{Aubert2018,Aubert2019,Aubert2020}, other outputs of the model are dimensioned by using the invariants of the parameter space path, particularly the stability of the magnetic Reynolds number and the QG-MAC force balance. The velocity field $\vecu$ and Alfvén velocity $\vecB/\sqrt{\rho\mu}$ are first expressed in a dimensionless manner in units of $\eta/D$, to respectively give local values of the magnetic Reynolds number $Rm$ and Lundquist number $S$. The results are then multiplied by the dimensional value of $\eta/D$ obtained from the choices made above. As we noted previously \citep{Aubert2018}, from $\vecB/\sqrt{\rho\mu}$ a dimensional value of $\vecB$ can be obtained, but this value is not realistic unless we are at the end of the path. At 71 percent of the path, this procedure yields $B=1.4\ut{mT}$ and the mismatch with the Earth estimate $B=4\ut{mT}$ \citep{Gillet2010} is no longer large, because the Alfvén time scale is now close to the Earth value (Table \ref{outputs1}b). For the purpose of comparing the output of our path models to the geomagnetic field, and following our conventions taken for other path positions \citep{Aubert2018,Aubert2019}, we simply adopt $B=4\ut{mT}$ for dimensioning the magnetic field amplitude in all models. Finally, to dimension the density anomaly field $C$ we take advantage of the preservation of the balance between buoyancy and Coriolis force along the path \citep{Aubert2020}. The dimensionless density anomaly field is therefore expressed in units of $\rho\Omega\eta/g_{o}D$ and the result is then multiplied with Earth's core dimensional estimate for $\rho\Omega\eta/g_{o}D$ obtained with our choices for $\eta$ and the other values provided in section \ref{intro}. 

\subsection{\label{methodseries}Time series extracted from the simulation}
The physical time span integrated for the 71p model is 117 core overturns $\tau_{U}$ (Table \ref{outputs1}a), which corresponds to 14000 years of physical time, using the rescaling procedures introduced in section \ref{doutputs}. It should be emphasized that achieving this run length while simulating strongly scale-separated dynamical features comes at a sizeable numerical cost. The numerical time step of the computation has indeed been capped to about a hundredth of the rotational time $\tau_{\Omega}$. This is needed because of the explicit treatment of the Coriolis force in the time stepping scheme. In the later part of the computation, the time step for instance corresponds to 0.3 hours of physical time. The 71p run required about 246 million numerical time steps and 15 million single-core CPU hours spent over the course of 2 years. These figures highlight that the 71p model is an extreme computational endeavour, despite the use of approximations to reduce its spatial complexity. To tackle this challenge, significant code optimisations have been performed during the computation, to reach a wall time per iteration of 0.06 seconds on 2560 cores of the AMD-Rome partition of GENCI-TGCC in France. 

A number of output time series are extracted from the model in addition to the time-averaged diagnostics introduced above. As time scales shorter than the numerical day $2\pi\tau_{\Omega}$ (11.8 days in the 71p model) contain a negligible amount of energy \citep{Aubert2018}, the temporal spacing between samples has been set to this value. Time series of the core surface magnetic field $\vecB(r_{o},\theta,\varphi,t)$ and velocity field $\vecu(r_{o},\theta,\varphi,t)$ have been systematically recorded up to spherical harmonic degree 30. \rev{As magnetic time series in our earlier models \citep{Aubert2018,Aubert2019} were only recorded up to degree 13, here we further truncate $\vecB(r_{o},\theta,\varphi,t)$ after degree 13 for consistency.} Time derivatives $\dot{\vecu}=\partial \vecu/\partial t$ (the flow acceleration), \rev{$\dot{\vecB}$ (the magnetic variation) and $\ddot{\vecB}$ (the magnetic acceleration)} are computed by finite-differencing. We define time series for the energies $E_\mathrm{SV}, E_\mathrm{SA}$ of the magnetic variation and acceleration as
\begin{eqnarray}
E_{\mathrm{SV}}(t)= \dfrac{1}{S} \int_{S} \dot{\vecB}^{2}(r_{o},\theta,\varphi,t)~\mathrm{d}S,\\
E_{\mathrm{SA}}(t)= \dfrac{1}{S} \int_{S} \ddot{\vecB}^{2}(r_{o},\theta,\varphi,t)~\mathrm{d}S.
\end{eqnarray}
Here $S$ is the spherical surface at $r=r_{o}$. These quantities can also be evaluated at Earth's surface $S_\mathrm{E}$ located at $r_\mathrm{E}=1.83r_{o}$, where $\vecB$ can be upward-continued by assuming an insulating mantle. As in our previous study \citep{Aubert2019}, an Earth surface jerk energy time series is defined in a way that factors in the limited resolution of geomagnetic observations, as a sliding energy difference of magnetic acceleration averaged over consecutive 3-years windows:
\begin{equation}
E_\mathrm{J} (t) = \dfrac{1}{S_\mathrm{E}} \int_{S_{\mathrm{E}}} \left(\left[\ddot{\vecB}\right]_{t}^{t+3 \ut{yr}}-\left[\ddot{\vecB}\right]_{t-3 \ut{yr}}^{t}\right)^{2}(r_{E},\theta,\varphi,t)~\mathrm{d}S,\label{EJdef}
\end{equation}
where the square brackets stand for time averaging. We define the energy associated to the core surface flow acceleration $\dot{\vecu}(\vc{r_{o}},t)$ up to spherical harmonic degree 30 as
\begin{equation}
K(t)=\dfrac{1}{S} \int_{S} \dot{\vecu}^{2}(r_{o},\theta,\varphi,t)~\mathrm{d}S.
\end{equation}
We label as $\dot{\vecu}^\mathrm{S}(r_{o},\theta,\varphi,t)$ the equator-symmetric component of $\dot{\vecu}$, with the associated energy $K^\mathrm{S}$. In the asymptotic regime reached beyond 29 percent of the path, the large-scale flow acceleration is dominantly equator-symmetric \citep{Aubert2018}, as confirmed here by ratios $K^\mathrm{S}/K$ increasing from 0.7 to 0.84 between 29 and 71 percent of the path. We further denote as $\dot{u}_{\varphi}^\mathrm{ZS}(r_{o},\theta,\varphi,t)$ the axisymmetric (zonal) average of $\dot{\vecu}^\mathrm{S}\cdot\vc{e}_{\varphi}$, and the corresponding energy as $K^\mathrm{ZS}$. By using a standard Thomson multitaper method with concentration half-bandwidth $W=4/\Delta t$ (where $\Delta t$ is the run length), we also decompose the Earth-surface magnetic accceleration $\ddot{\vecB}(r_{E},\theta,\varphi)$, the core surface flow accelerations $\dot{\vecu}(r_{o},\theta,\varphi)$ and $\dot{u}_{\varphi}^\mathrm{ZS}(r_{o},\theta,\varphi)$ in the frequency domain to calculate the respective energy density spectra $\hat{E}_\mathrm{SA}(f)$, $\hat{K}(f)$ and $\hat{K}^{\mathrm{ZS}}(f)$. Given the duration of our runs (see Table \ref{inputs}) this ensures that the overturn frequency $1/\tau_{U}$ is well resolved in all model cases. 

For reasons of storage space, full three-dimensional outputs of the simulation state at native resolution are available only in selected portions of the numerical computation. In these portions and as in \cite{Aubert2018}, we introduce the \rev{azimuthal} acceleration on geostrophic cylinders (whose axis is aligned with the rotation axis $\vc{e}_{z}$):
\begin{equation}
\dot{u}_{g}(s,t)= \dfrac{1}{2\pi(z_{+}-z_{-})}\int_{0}^{2\pi}\int_{z_{-}}^{z_{+}} \dot{\vecu}(s,\varphi,z,t)\cdot\vc{e}_{\varphi}\, \mathrm{d} z \mathrm{d}\varphi,\label{ug}
\end{equation}
\rev{as well as the Alfvén speed given by the part of poloidal magnetic field permeating these cylinders:}
\begin{equation}
c_{A}^{2}(s,t) = \dfrac{1}{2\pi(z_{+}-z_{-})\rho\mu}\int_{0}^{2\pi}\int_{z_{-}}^{z_{+}} (\vecB(s,\varphi,z,t)\cdot\vc{e}_s)^2\, \mathrm{d} z \mathrm{d}\varphi.\label{ca}
\end{equation}
Here $z_{-}$ \rev{and} $z_{+}$ are the ordinates along $\vc{e}_{z}$ of the intersections between geostrophic cylinders of radius $s$ and the spherical boundaries of the shell. 

\section{\label{results}Results}
\subsection{\label{pathupdate}The 71p model versus earlier solutions along the parameter space path}
\begin{figure}
\centerline{\includegraphics[width=9cm]{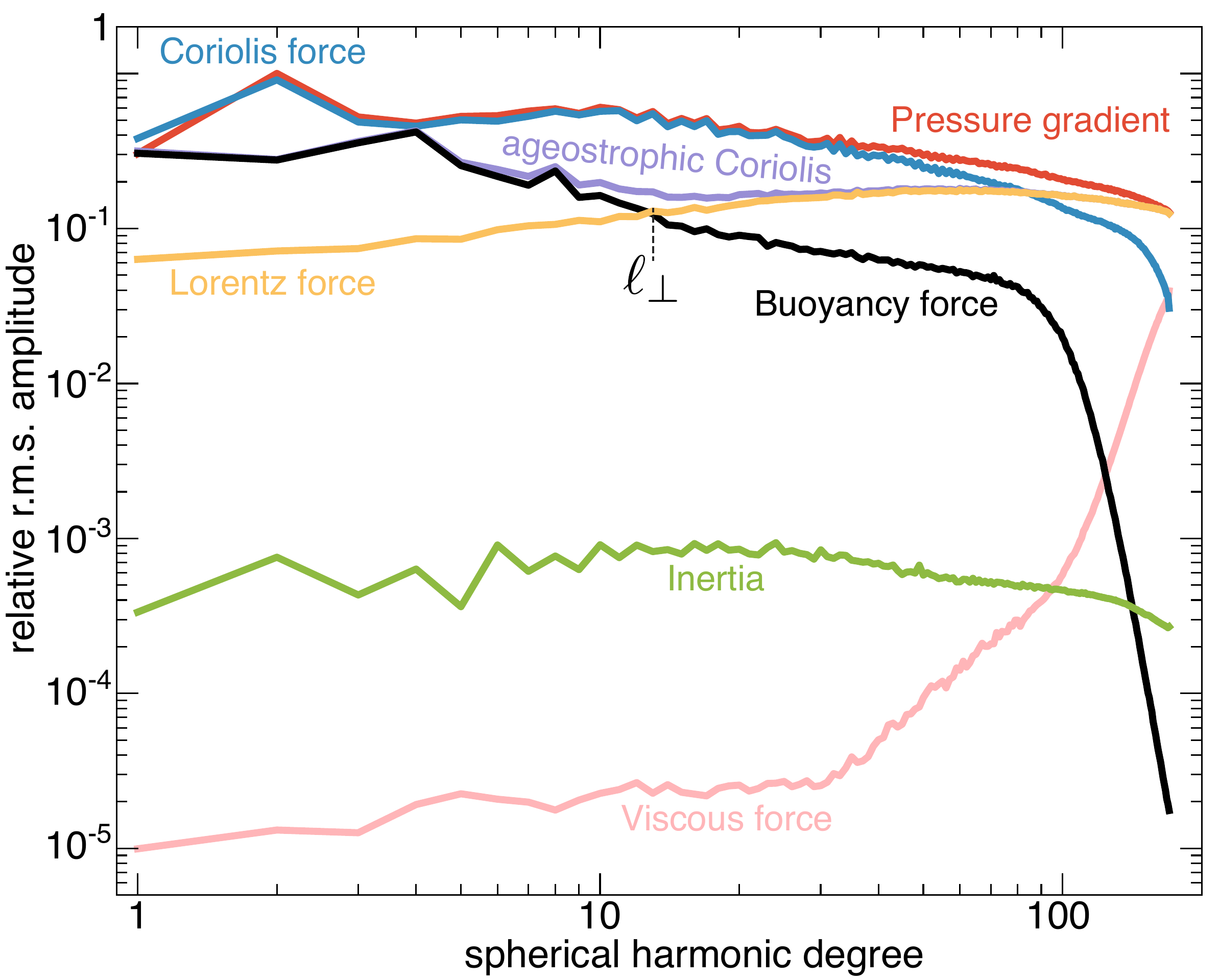}}
\caption{\label{forcebal} Root-mean-squared amplitude of the forces in a snapshot of the 71p model, presented as functions of the spherical harmonic degree $\ell$ (see section \ref{pathcomp}). All forces are normalised respectively to the maximum reached by the pressure gradient. The snapshot was taken from the later part of the run where $\ell_\mathrm{max}=170$ and $q_{h}=1.09$ (see Fig. \ref{SVSAmorph}).}
\end{figure}

Our first task is to update our previous analysis of the path in parameter space \citep{Aubert2017,Aubert2018,Aubert2019} in the light of the new data point acquired at 71 percent of this path. Figure \ref{forcebal} presents a scale-dependent force balance diagram obtained with a snapshot of the 71p model. Similarly to our previous results at earlier positions along the path \citep{Aubert2017,Aubert2018,Aubert2019}, a leading-order QG balance is still observed between the pressure and Coriolis forces. At the next order follows a MAC balance between the buoyancy, Lorentz and ageostrophic Coriolis forces. The cross-over between the Lorentz and buoyancy forces defines the harmonic degree $\ell_{\perp}=\pi D/d_{\perp}$ where the triple balance is exactly respected, with a value $\ell_{\perp}=13$ that has only slightly evolved since Midpath conditions where $\ell_{\perp}=12$ \citep{Aubert2019b}. This scale, at which convective energy is injected into the magnetic field, is therefore confirmed to remain stable along the path. At the second order in amplitude, inertial forces are now located two orders of magnitude below the MAC forces. As previously shown in \cite{Schwaiger2019}, the widening of the gap between MAC and inertial forces along the path scales with the inverse squared Alfvén number $A^{-2}$, in line with the evolution observed between Midpath \citep[see Fig. 1 in][]{Aubert2018} and 71 percent of path conditions. In the 71p model, viscosity comes five order of magnitude below the Coriolis force at large scales. Our choice for spatial resolution and strength of hyperdiffusivity in the later part of the run (from where the snapshot was taken) result in a negligible influence of viscosity up to spherical harmonic $\ell\approx 80$, after which its strength ramps up and exceeds that of inertia, as was previously the case in the Midpath model.

\begin{figure}
\centerline{\includegraphics[width=14cm]{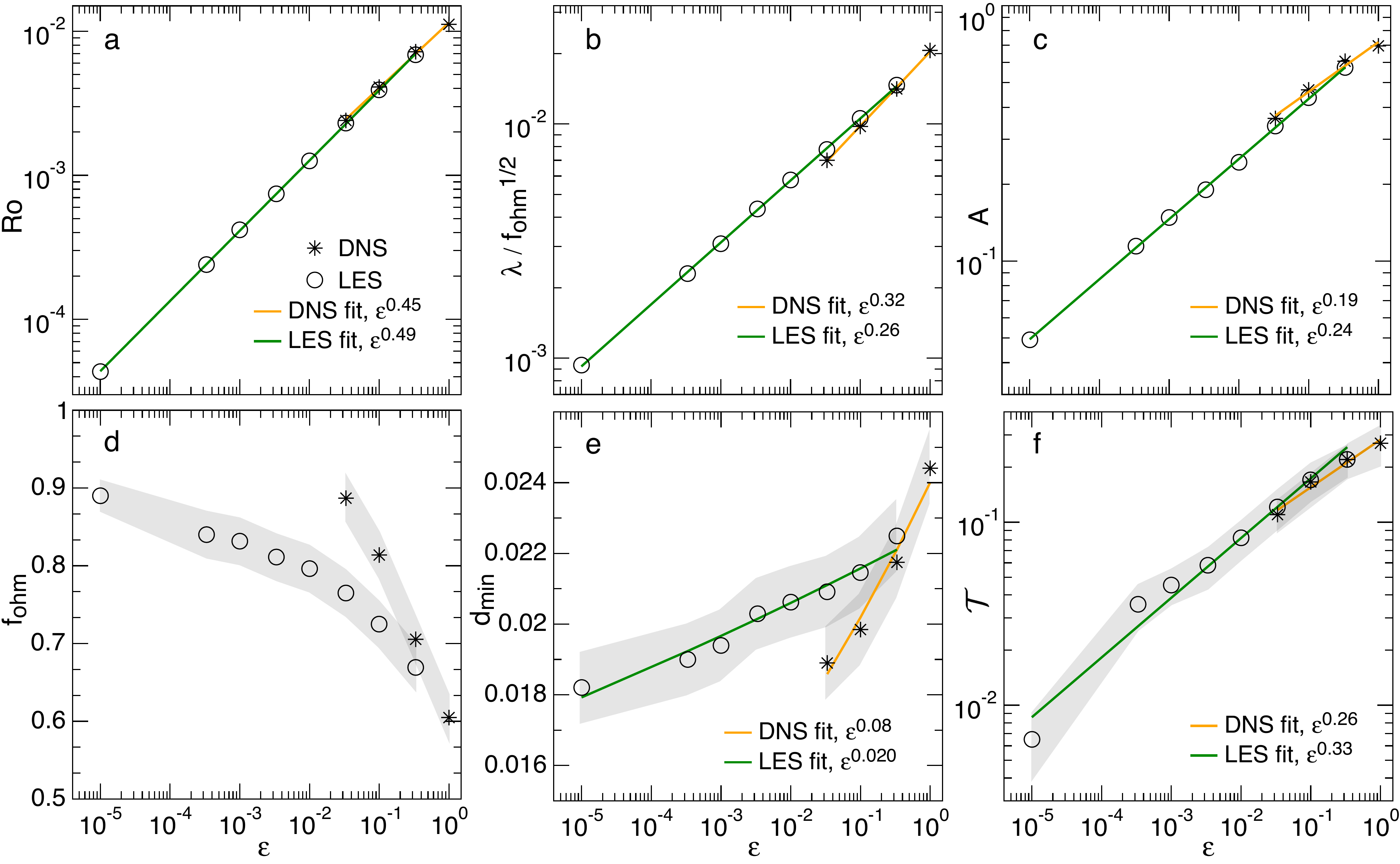}}
\caption{\label{scalings} Evolution of the Rossby number $Ro$ (a), Lehnert number $\lambda$ (b), Alfvén number $A$ (c), Ohmic fraction $f_\mathrm{ohm}$ (d), magnetic dissipation length scale $d_\mathrm{min}$ (e) and Taylor constraint enforcement level $\mathcal{T}$ (f) along the parameter space path, as functions of the path parameter $\epsilon$ (Earth's core conditions are towards the left of the graphs). Shown are results from hyperdiffusive solutions along the path (LES or large-eddy-simulations) up to 71 percent of the path ($\epsilon=\te{-5}$), and the direct numerical simulations (DNS) up to 21 percent of the path ($\epsilon=3.33~\te{-2}$). See Tables \ref{outputs1}a and \ref{outputs2} for data between 29 and 71 percent of the path, and \cite{Aubert2017,Aubert2019b} for other numerical values. Green and orange lines respectively show the results of least-squares power-law fits to the LES and DNS data, which closely adhere to the theories of \cite{Aubert2017,Davidson2013}, respectively. Shaded grey areas represent the $\pm 1$ standard deviation, where applicable.}
\end{figure}

The stability of the force balance implies that previously determined scaling laws for the main model outputs are essentially unchanged when adding the 71p model data. We have previously shown \citep{Aubert2017} that along the path \rev{and for the hyperdiffusive, large-eddy (LES) simulations}, the QG-MAC balance, together with the constancy of the magnetic Reynolds number $Rm$ imply the following scaling laws for the Rossby number and ohmic fraction-corrected Lehnert number:
\begin{eqnarray}
Ro \sim \epsilon^{1/2},\\
\lambda/f_\mathrm{ohm}^{1/2} \sim \epsilon^{1/4}.
\end{eqnarray}
The way the path models adhere to these scalings is essentially unchanged when adding the 71p model data points (Fig. \ref{scalings}a,b). From this, the scaling $A=Ro/\lambda \sim \epsilon^{1/4}$ is also respected (Fig. \ref{scalings}c). Relative to the Midpath model ($\epsilon=3.33~\te{-4}$), the Ohmic fraction has increased at 71 percent of the path to reach $f_\mathrm{ohm}=0.89$ (Fig. \ref{scalings}d), meaning that despite the use of hyperdiffusivity, dissipation is essentially of Ohmic nature in this model. The magnetic dissipation length scale $d_\mathrm{min}$ remains essentially invariant along the path  (Fig. \ref{scalings}e, note the linear ordinate axis), a consequence of the constant magnetic Reynolds number and constrained spatial structure of the solution. Finally, the level to which the solution approaches a Taylor state \citep{Taylor1963} can be measured by the Taylorisation level $\mathcal{T}$ (Fig. \ref{scalings}f), with the 71p model exhibiting the highest levels of taylorisation observed to date along the path (lowest value of $\mathcal{T}$). Somewhat surprisingly, this model appears to respect the Taylor constraint even more strongly than the expectation that could be drawn from the previous models. For reference, the scaling results from fully resolved direct numerical simulations (DNS) are also reported on Fig. \ref{scalings}. These simulations have so far been computed only down to 21 percent of the path \citep[$\epsilon=3.33~\te{-2}$,][]{Aubert2019b}. In this previous study, we have shown that these adhere to the power laws predicted by the QG-MAC theory \citep{Davidson2013,Aubert2017} i.e. $Ro \sim \epsilon^{4/9}$, $\lambda/f_\mathrm{ohm}^{1/2} \sim \epsilon^{1/3}$, $d_\mathrm{min}\sim\epsilon^{1/12}$. The difference between path and DNS theories scalings essentially stems from the fact that the length scales are constrained by hyperdiffusivity in the former situation, while they are weakly evolving in the latter case. 

\begin{figure}
\centerline{\includegraphics[width=14cm]{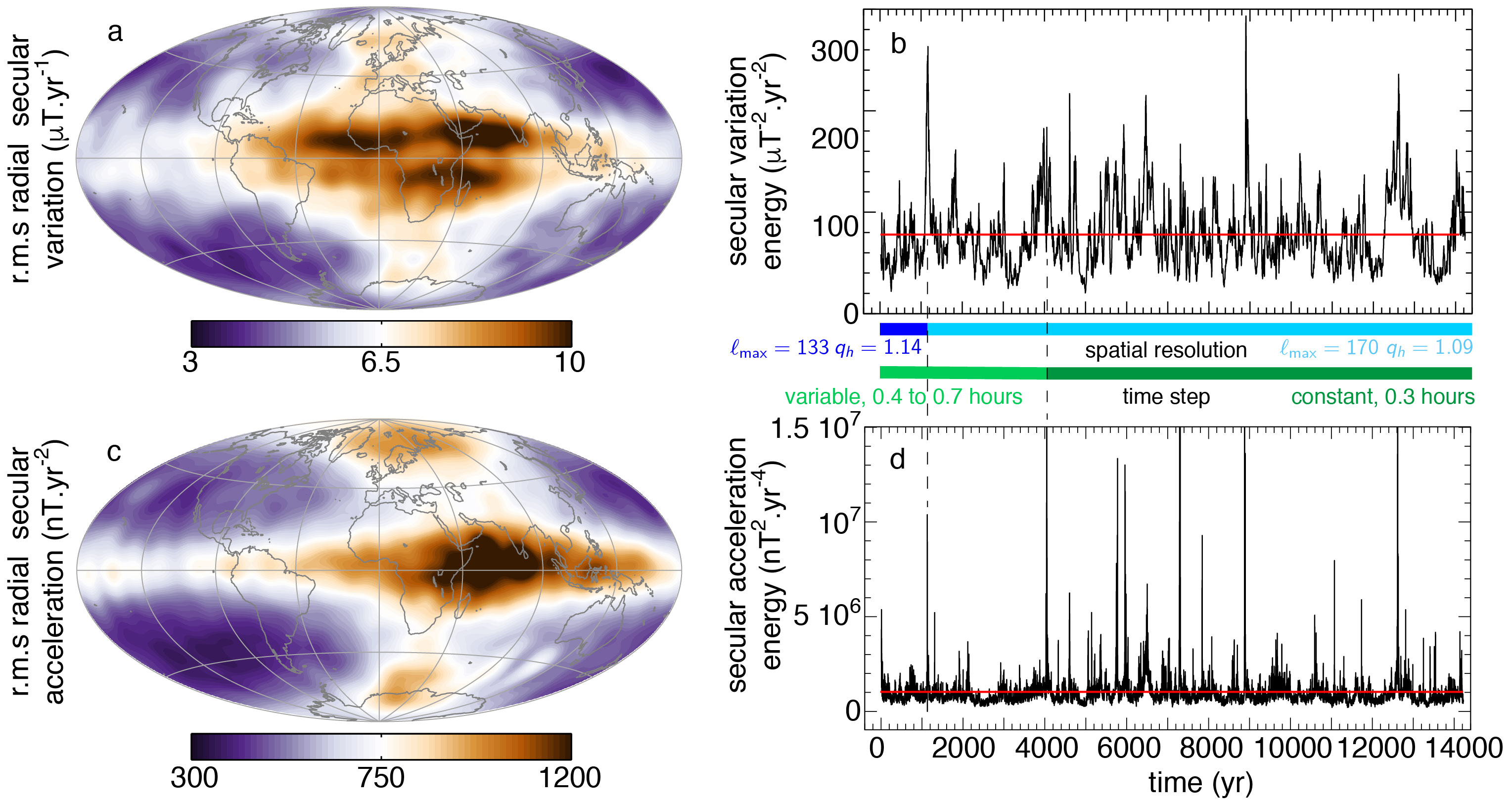}}
\caption{\label{SVSAmorph} Magnetic variation and acceleration properties of the 71p model. a,c: Hammer projections of the \rev{time-averaged}, root-mean squared radial magnetic variation $\dot{\vecB}\cdot\vc{e}_{r}$ (a) and acceleration $\ddot{\vecB}\cdot\vc{e}_{r}$ (c) at the core surface, from magnetic field output up to spherical harmonic degree 13. b,d: Time series of the magnetic variation and acceleration energies $E_\mathrm{SV}$ (b) and $E_\mathrm{SA}$ (d) at the core surface. Red lines in (b,d) mark the time-averaged values $[E_\mathrm{SV}]$ and $[E_\mathrm{SA}]$. The blue and green bands between (b) and (d) locate the notable events related to spatial and temporal resolution changes in the model computation.}
\end{figure}

In Fig. \ref{SVSAmorph} we next turn to key aspects of magnetic variation and acceleration in the 71p model, which we analyse in a manner similar to Fig. 3 of \cite{Aubert2018}. Consistently with the defining principles of the path, and with the approximately constant values of $\tau_{U},\tau_\mathrm{SV}^{1}$ observed along this path (Table \ref{outputs1}), the properties of magnetic variation at 71 percent of the path (Fig. \ref{SVSAmorph}a,b) remain essentially similar to those observed since the start of the path, with an Earth-like geographic localisation that is essentially dictated by the \rev{heterogeneous mass anomaly flux imposed at the inner boundary} \citep{Aubert2013b}. The pattern of magnetic acceleration is also similar to that previously obtained from 29 percent of the path onwards, with this acceleration being mainly observed within an equatorial band in the Eastern hemisphere ($0^\mathrm{o}\mathrm{E}-180^\mathrm{o}\mathrm{E}$). The 71p model confirms that the relative dominance of equatorial over polar magnetic acceleration continues to increase as we advance along the path. As was the case in previous models, the 71p model features intermittent pulses in the magnetic acceleration energy, with a clear increase in strength and frequency relative to the Midpath conditions. Relatively to the midpath Model, the time-averaged acceleration energy $[E_\mathrm{SA}]$ of the 71p model (red line in Fig. \ref{SVSAmorph}d) consequently increases by about 25 percent, while $[E_\mathrm{SV}]$ is almost unchanged (Fig. \ref{SVSAmorph}b), leading to a time scale $\tau_\mathrm{SA}^{0}\approx \sqrt{[E_\mathrm{SV}]/[E_\mathrm{SA}]}=10.6\ut{yr}$ slightly shorter that its Midpath value $11.8\ut{yr}$ (Table \ref{outputs1}b). The general picture provided by the magnetic variation and acceleration in the 71p model is therefore consistent with the previously described evolution along the path. The kinematics and convectively-driven part of the dynamics remain invariant, and the dynamics is also gradually enriched in short-term, intermittent acceleration pulses.

\begin{figure}
\centerline{\includegraphics[width=14cm]{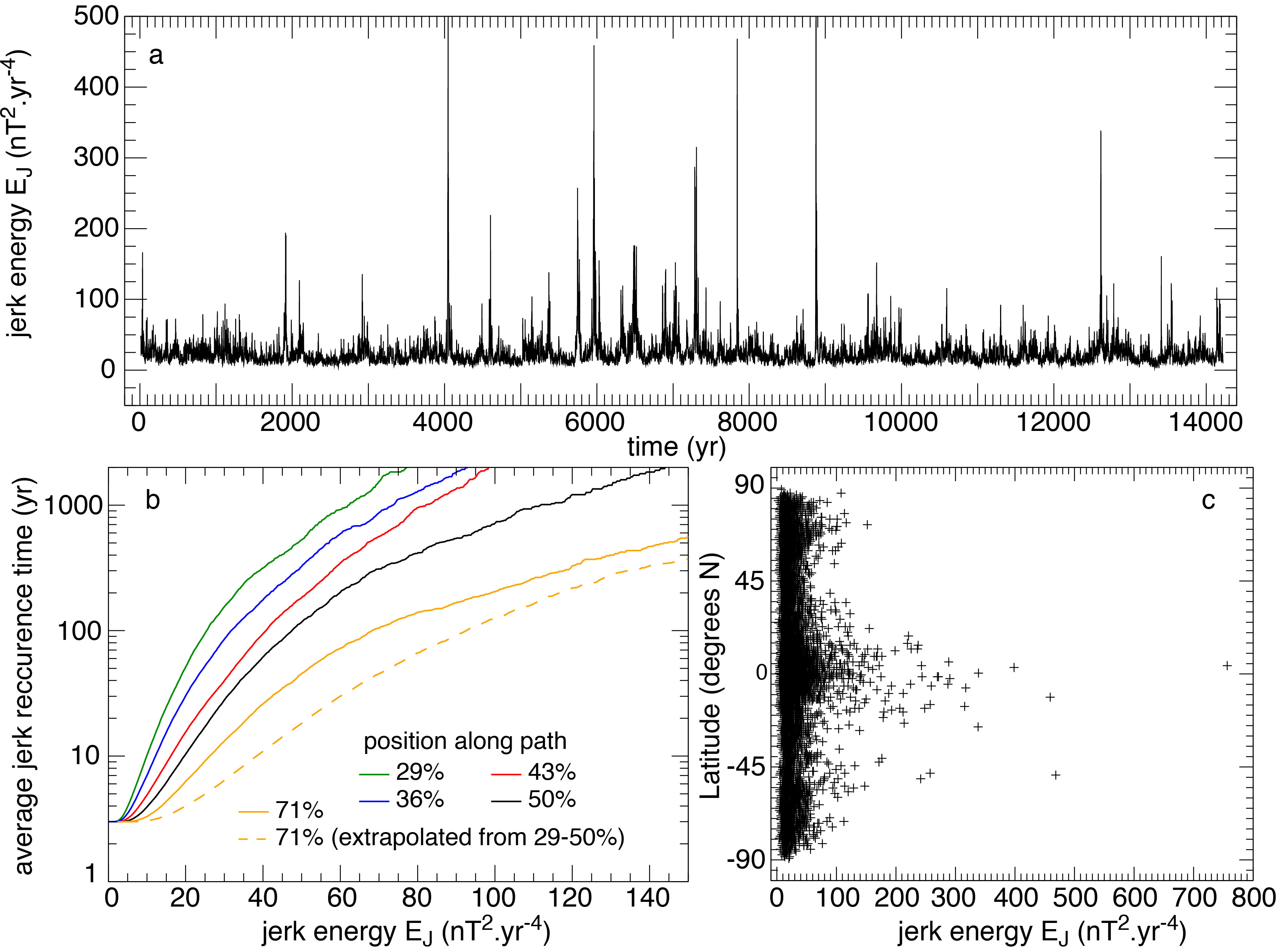}}
\caption{\label{EJ} a: Time series of the Earth surface jerk energy $E_\mathrm{J}$ (see section \ref{methodseries}). b: Distribution of jerk energy as a function of the jerk recurrence time, for models previously computed in the range 29 percent to 50 percent of the parameter space path \citep{Aubert2019} and the 71p model. Shown in dashed line is the expected distribution at 71 percent of the path from an extrapolation $E_\mathrm{J}\sim \epsilon^{-0.19}$ previously obtained from the behaviours between 29 and 50 percent of the path \citep{Aubert2019}. c: Distribution of jerk energy in the 71p model (sampled each year) as a function of the latitude of the amplitude maxima observed in the radial magnetic acceleration at the core surface. Jerk statistics in b,c are computed in the time window  $[4200\ut{yr},14200\ut{yr}]$ where the spatio-temporal resolution of the computation is optimal (see coloured bands under Fig. \ref{SVSAmorph}b). This interval also avoids two spurious acceleration pulses near times 1130 and 4180 years, corresponding to resolution changes.}
\end{figure}

The magnetic acceleration pulses have been previously related to the dynamics and emergence of non-axisymmetric Alfvén waves at the core surface, on a rapid time scale commensurate with $\tau_{A}$. This leads to geomagnetic signatures that reproduce the characteristics of recent geomagnetic jerks \citep{Aubert2019}. To characterise the evolving properties of magnetic acceleration pulses and jerks along the parameter space path, we previously found useful to analyse time series of the jerk energy $E_\mathrm{J}$ (Fig. \ref{EJ}a and equation  \ref{EJdef}). This quantity mirrors the occurence of acceleration pulses in a way that is more straightforwardly comparable to satellite-based determinations of the geomagnetic acceleration. New to the 71p model is the occurrence of extremely fast magnetic acceleration pulses (for instance at times 11070 and 11740 in Fig. \ref{SVSAmorph}d) that are smoothed by the 3-year averaging process involved in the definition of $E_\mathrm{J}$ (see corresponding times in Fig. \ref{EJ}a). Indeed, while this observation-based averaging window was clearly shorter than the typical Alfvén time over which the jerks develop in the models at 29 to 50 percent of the path, in the 71p model it is now only slightly shorter than $\tau_{A}=5.8\ut{yr}$. Jerk recurrence statistics are computed in Fig. \ref{EJ}b as in \cite{Aubert2019}, by dividing the duration of the 71p model run with the number of samples reaching or exceeding a given value of $E_\mathrm{J}$ and representing the resulting estimates for jerk recurrence time as a function of $E_\mathrm{J}$. This representation confirms that the 71p model produces significantly more energetic jerks than the Midpath model at any recurrence time. However, the energy levels that are obtained are slightly recessed relative to the extrapolation that could be made in \cite{Aubert2019} from models spanning 29 to 50 percent of the path. This obviously relates to the smoothing effect described above. It is finally useful to analyse the distribution of $E_\mathrm{J}$ with the latitude of maxima in the associated magnetic acceleration pulses at the core surface (Fig. \ref{EJ}c). Consistently with the root-mean-squared pattern seen in Fig. \ref{SVSAmorph}c, magnetic acceleration pulses occur preferably at high latitudes (near the projection on the core-mantle boundary of the geostrophic cylinder tangent to the inner core, hereafter called the tangent cylinder) and the equator, with equatorial dominance for the strongest events. \rev{The residual equatorial asymmetry observed in Fig. \ref{EJ}c for strong events links with the rarity of such events and the limited duration of the sequence.} In the 71p model, about 80 percent of the events of magnitude $E_\mathrm{J}\ge 100 \ut{nT^{2}/yr^{4}}$ have core surface foci at latitudes less than 20 degrees away from the equator. The 71p model therefore pursues the trend previously reported in \cite{Aubert2018} towards equatorial localisation of pulses and jerks as we advance along the path. 
 
\subsection{\label{interplay}Frequency ranges for waves and convection}

\begin{figure}
\centerline{\includegraphics[width=14cm]{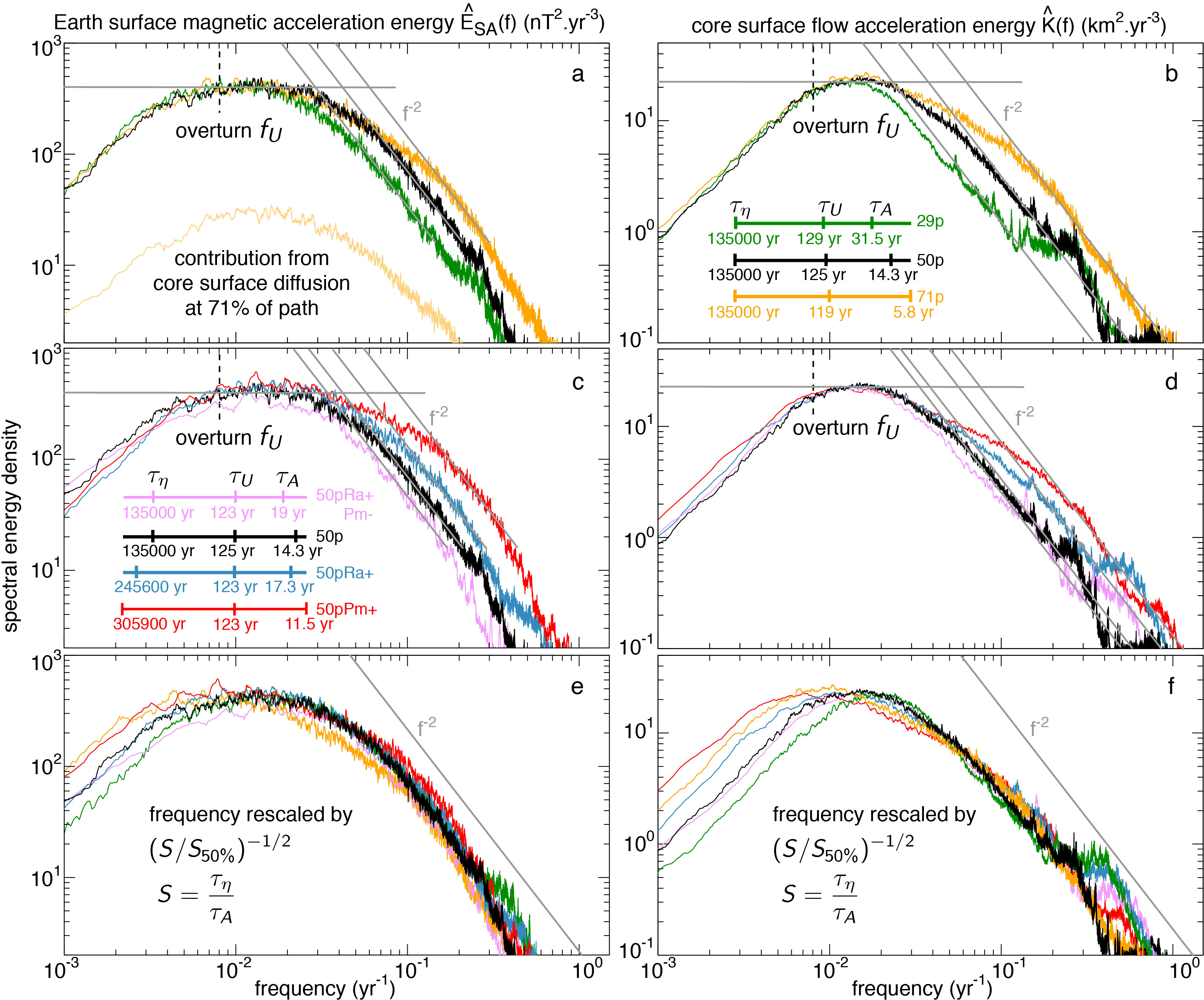}}
\caption{\label{psd} Frequency domain spectral density $\hat{E}_\mathrm{SA}(f)$ of the magnetic acceleration energy  at Earth's surface (a,c,e), and $\hat{K}(f)$ of the flow acceleration energy  at the core surface (b,d,f), for models taken along (a,b) and across (c,d)  the parameter space path. a,b: spectra of the 29 percent, 50 percent and 71 percent of path models, plotted together in a. with the additional contribution of core surface magnetic diffusion in the 71p model. b,d: spectra of the 50 percent of path, 50pRa+Pm-, 50pRa+ and 50pPm+ models. In a-d the dashed vertical line marks the overturn frequency $f_{U}=1/\tau_{U}$. Flat and slanted grey lines respectively mark the power-laws $f^{0}$ and $f^{-2}$\rev{, the crossing of which defines the cut-off frequency $f_{c}$}. The relative positions and numerical values of the time scales $\tau_{\eta}$, $\tau_{U}$ and $\tau_{A}$ are also recalled. e,f: respective compounds of a,c and b,d, with frequency axis rescaled for each model by $(S/S_{50\%})^{-1/2}$, where $S=\tau_{\eta}/\tau_{A}$ and $S_{50\%}$ is the Lundquist number of the Midpath model.}
\end{figure}

In Fig. \ref{psd} we present the energy density spectra $\hat{E}_\mathrm{SA}(f)$ and $\hat{K}(f)$ of the Earth surface magnetic acceleration energy and core surface flow acceleration energy for models located between 29 and 71 percent of the path as well as across the path at 50 percent. At frequencies up to that of the overturn, $f_{U}=1/\tau_{U}=U/D$, Fig. \ref{psd}a-d confirms that all models feature similar spectral energy densities, an indication of the invariant dynamics caused by slow secular convection. At frequencies beyond $f_{U}$, all models also feature a range where the energy density profile is flattened. Considering that each successive time derivative multiplies the energy density by $f^{2}$, this flat range corresponds to a $f^{-2}$ range in the energy spectral density of the observed geomagnetic variation, and to a $f^{-4}$ range in that of the magnetic energy, both of which have received reasonable observational support \citep{DeSantis2003,Lesur2018}. The flat range for flow acceleration corresponds to a $f^{-2}$ range for flow velocity, which has previously been assumed as part of stochastic core flow modelling strategies \citep{Gillet2015,Gillet2019}. In the vicinity of the overturn frequency $f_{U}$, it is logical to associate this range of approximately flat spectra energy density to the signature of convective motions. At higher frequencies towards the decadal range, decay spectral tails following a $f^{-2}$ power law are observed both in flow and magnetic acceleration energies. \rev{A cut-off frequency $f_{c}$ that marks the start of the decay range is defined from the crossing between the flat ($f^{0}$) and slanted ($f^{-2}$) grey lines in Fig. \ref{psd}a-d. The decay spectral tails} can no longer be associated with convection because they are not invariant as we progress along the path (Fig. \ref{psd}a,b), as now better illustrated with the new data brought by the 71p model \rev{and the associated increase of the cut-off frequency $f_{c}$}. The combination of models along (Fig. \ref{psd}a,b) and across the path (Fig. \ref{psd}c,d) enables us to pinpoint the main control on $f_{c}$ as well as on the energy density within the decay range. The influence of the rotational time scale $\tau_{\Omega}$ can be immediately ruled out because the spectra of cross-path models with constant $Ro$ or $\lambda$ (see table \ref{outputs1}a) do not superimpose. The key time scale therefore appears to be the Alfvén time $\tau_{A}$, but the question remains whether it should be considered in proportion of the overturn time $\tau_{U}$ (i.e by using the Alfvén number $A$ as a control) or magnetic diffusion time $\tau_{\eta}$ (i.e by using $S$). 

\rev{A least-squares fit performed on the $f_{c}$ values extracted from Figs. \ref{psd}a-d yields $f_{c}\sim S^{0.5\pm 0.05}$, indicating that} all spectra can be collapsed by rescaling the frequency axis in proportion of $S^{-1/2}$, \rev{as further verified in} Fig. \ref{psd}e,f. \rev{The alternative scaling $f_{c}\sim A^{-1/2}$ can be readily discarded as the Alfvén numbers for the two cases 50pRa+ and 50pPm+ with magnetic Reynolds numbers $Rm=\tau_{\eta}/\tau_{U}\ge 2000$ obviously fail to compensate the shift with the other cases at $Rm\approx 1000$ (see table \ref{outputs1}a).} Given that the overturn frequency $f_{U}$ is constant prior to re-scaling the frequency axis, the scaling for the cut-off frequency hence writes 
\begin{equation}
f_{c} \approx f_{U} (S/S_{0})^{1/2},\label{cutoff}
\end{equation}
with $S_{0}\approx 900$ from our results. The cut-off frequency $f_{c}$ and $f^{-2}$ decay tails are similar for both the flow and magnetic acceleration energy spectra. In the decay range $f\ge f_{c}$, the functional form of these energy densities is therefore
\begin{eqnarray}
\hat{E}_\mathrm{SA}(f) \approx E_\mathrm{SA}^\mathrm{max} \left(\frac{f}{f_{c}}\right)^{-2} \approx \dfrac{E_\mathrm{SA}^\mathrm{max}f_{U}^{2}}{S_{0}} ~ S f^{-2},\label{Mscal}\\
\hat{K}(f) \approx K^\mathrm{max} \left(\frac{f}{f_{c}}\right)^{-2} \approx \dfrac{K^\mathrm{max}f_{U}^{2}}{S_{0}} ~ S f^{-2}\label{Kscal}.
\end{eqnarray}
Here $E_\mathrm{SA}^\mathrm{max}, K^\mathrm{max}$ are the values of the plateaus reached in the flat energy density range. Within the decay tails, the flow and magnetic energy densities therefore linearly increase with the number $S$ of Alfvén wave periods contained within a magnetic diffusion time. This strongly suggests that these decay tails $f\ge f_{c}$ are dominated by the contribution from Alfvén waves. The range $f\le f_{U}$ is the convective range, and the interval $[f_{U},f_{c}]$ where energy density presents a plateau is the range of interplay between waves and convection. As the Lundquist number $S$ increases, the plateau broadens because of the elevated contribution from Alfvén waves at its rightmost edge\rev{, with $S_{0}$ in equation (\ref{cutoff}) representing the minimal Lundquist number needed for this effect to occur, or more generally for waves to be significant in the simulation.} Writing the time-derivative of the magnetic induction equation 
\begin{equation}
\ddot{\vecB}=\nabla\times\left(\dot{\vecu}\times \vecB\right)+\nabla\times\left(\vecu\times \dot{\vecB}\right)+\eta\boldsymbol{\Delta} \dot{\vecB},\label{MISA}
\end{equation}
the similar spectral shapes (\ref{Mscal},\ref{Kscal}) also indicate that in the wave range at $f\ge f_{c}$, the first term in the right-hand-side dominates the production of magnetic acceleration, as previously anticipated for rapid dynamics \citep{Lesur2010b,Christensen2012,Aubert2018}. A direct calculation of the contribution by core surface magnetic diffusion to $\hat{E}_\mathrm{SA}(f)$ (Fig. \ref{psd}a) confirms that the term 
$\eta\boldsymbol{\Delta} \dot{\vecB}$ in (\ref{MISA}) is subdominant. The diffusive effects observed through the control of $S$ on $\hat{E}_\mathrm{SA}(f)$ and $\hat{K}(f)$ therefore originate in the bulk of the core. Incidentally, the functional forms (\ref{Mscal},\ref{Kscal}) may also be rewritten by involving the ratio $[(U/f) / \delta]^{2}$, where $U/f$ is a length scale excited by convection at frequency $f$ and $\delta=\sqrt{\eta \tau_{A}}$ is the bulk magnetic diffusion length scale at the Alfvén time. 

\subsection{Axisymmetric torsional Alfvén waves.}
\begin{figure}
\centerline{\includegraphics[width=14cm]{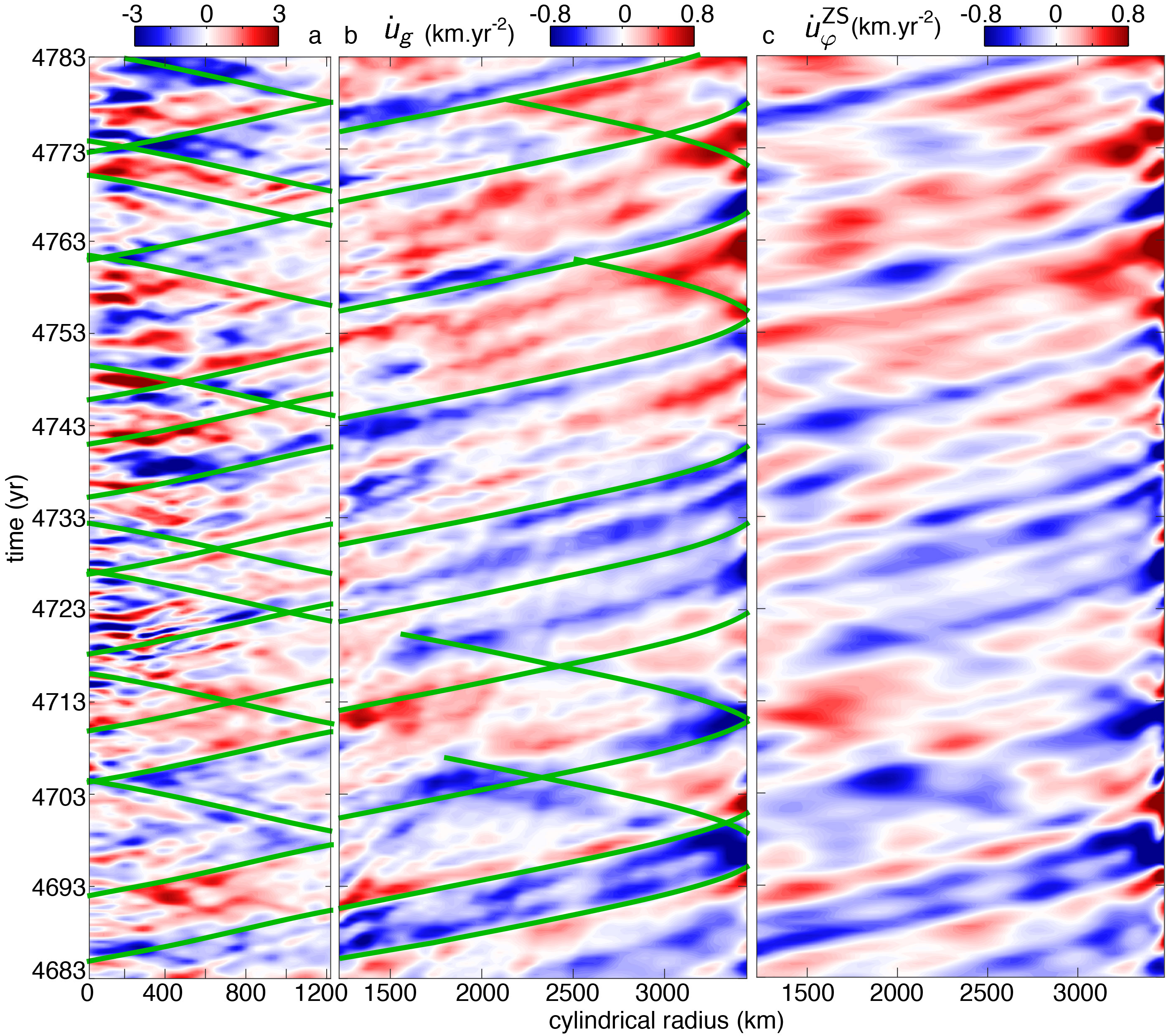}}
\caption{\label{TW71}Time-cylindrical radius diagrams of the geostrophic flow acceleration $\dot{u}_{g}$ (equation \ref{ug}) sampling the interior of the fluid in the Northern part of the tangent cylinder (a), and the region outside the tangent cylinder (b), at native resolution in the 71p model. c: time-cylindrical radius diagram obtained at the core surface and outside the tangent cylinder by representing the equator-symmetric, axisymmetric azimuthal flow acceleration $\dot{u}_{\varphi}^\mathrm{ZS}$ up to spherical harmonic degree 30. Note the larger extremes of the color map in a. relatively to b,c. Green curves in a.,b. correspond to outwards and inwards ray-tracing theoretical propagation tracks at the column-averaged Alfvén speed $c_{A}(s,t)$ (equation \ref{ca}).}
\end{figure}

We can pursue our analysis by examining either the flow or magnetic acceleration signal carried by the waves. The former is more convenient, because rapid azimuthal flow acceleration presents a highly columnar (and therefore equator-symmetric) structure at advanced positions along the path. Torsional waves are a straightforward and interesting case study of the Alfvén waves present in the system, because they can be easily isolated in the axially-columnar, axisymmetric, azimuthal part $\dot{u}_{g}$ of $\dot{\vecu}$ defined by equation (\ref{ug}). Fig. \ref{TW71}a,b shows time-cylindrical radius representation of $\dot{u}_{g}$ during a 100-yr sequence of the 71p model. As in Fig. 10 of \cite{Aubert2018}, the presence and dominance of torsional waves in this signal is attested using the similarity of the patterns with ray-tracing tracks that represent propagation at the theoretical speed $c_A(s,t)$ (equation \ref{ca}) commensurate with the one-dimensional Alfvén velocity $D/\sqrt{3}\tau_{A}\approx 225 \ut{km/yr}$. Compared to Fig. 10 of \cite{Aubert2018}, the signal outside the tangent cylinder (Fig. \ref{TW71}b) has logically increased in propagation speed (because of the decrease of $\tau_{A}$ from Midpath to 71p conditions), but less trivially also in amplitude. The waves in the northern part of the tangent cylinder (Fig. \ref{TW71}a) are even faster (as they sample a stronger magnetic field), and also more intense. \rev{As the model features a jump in $c_{A}$ of about 15 percent across the tangent cylinder at $s=1220 \ut{km}$, similarly to \cite{Teed2014} we mainly observe wave transmission across this interface, in agreement with classical electromagnetic continuity relationships \citep[e.g.][]{Alfven1963}. Wave excitation and reflection events at the tangent cylinder are less obvious but can also be seen sporadically. In contrast, reflected patterns at the core-mantle boundary are clearer in Fig. \ref{TW71}b, and qualitatively suggest an increase of the reflection coefficient relative to Midpath conditions \citep[Fig. 10 of][]{Aubert2018}, where reflection was elusive. Given that our simulations operate with stress-free boundary conditions, this results appears to be at variance with the plane layer theory laid out in \cite{Schaeffer2012,Schaeffer2016,Gillet2017}, which predicts that the reflection coefficient should not depend on the imput parameters varied along the path (most notably $Pm$), and should even decrease due to the presence of a conducting outer layer and the increasing Lundquist number $S$. We anticipate that the discrepancy ties with the treatment of the singularity present for torsional waves at the equator of the core-mantle boundary, which is probably affected by hyperdiffusivity as the wave pattern shrinks down to small latitudinal length scales. The 71p model is less affected by this problem than the Midpath model as it operates with higher native resolution and reduced hyperdiffusivity (see table \ref{inputs}). With stress-free boundaries and the conducting layer, the plane layer theory predicts a reflection coefficient $R=(1-Q)/(1+Q)$ \citep{Schaeffer2016}, where the quality factor writes $Q= S (\Delta \sigma_{m}/ D \sigma) (B_r (r_{o})/B)$. The conductance ratio is $\Delta \sigma_{m}/ D \sigma=\te{-4}$ for all models along the path, and the ratio $B_r (r_{o})/B\approx 1/8$ of the radial magnetic field at the core surface to internal magnetic field is also approximately constant, as a consequence of static invariance of the solution. For the 71p model this theory therefore predicts $Q=0.3$ and $R=0.5$, qualitatively in line with Fig. \ref{TW71}b.}

\begin{figure}
\centerline{\includegraphics[width=8cm]{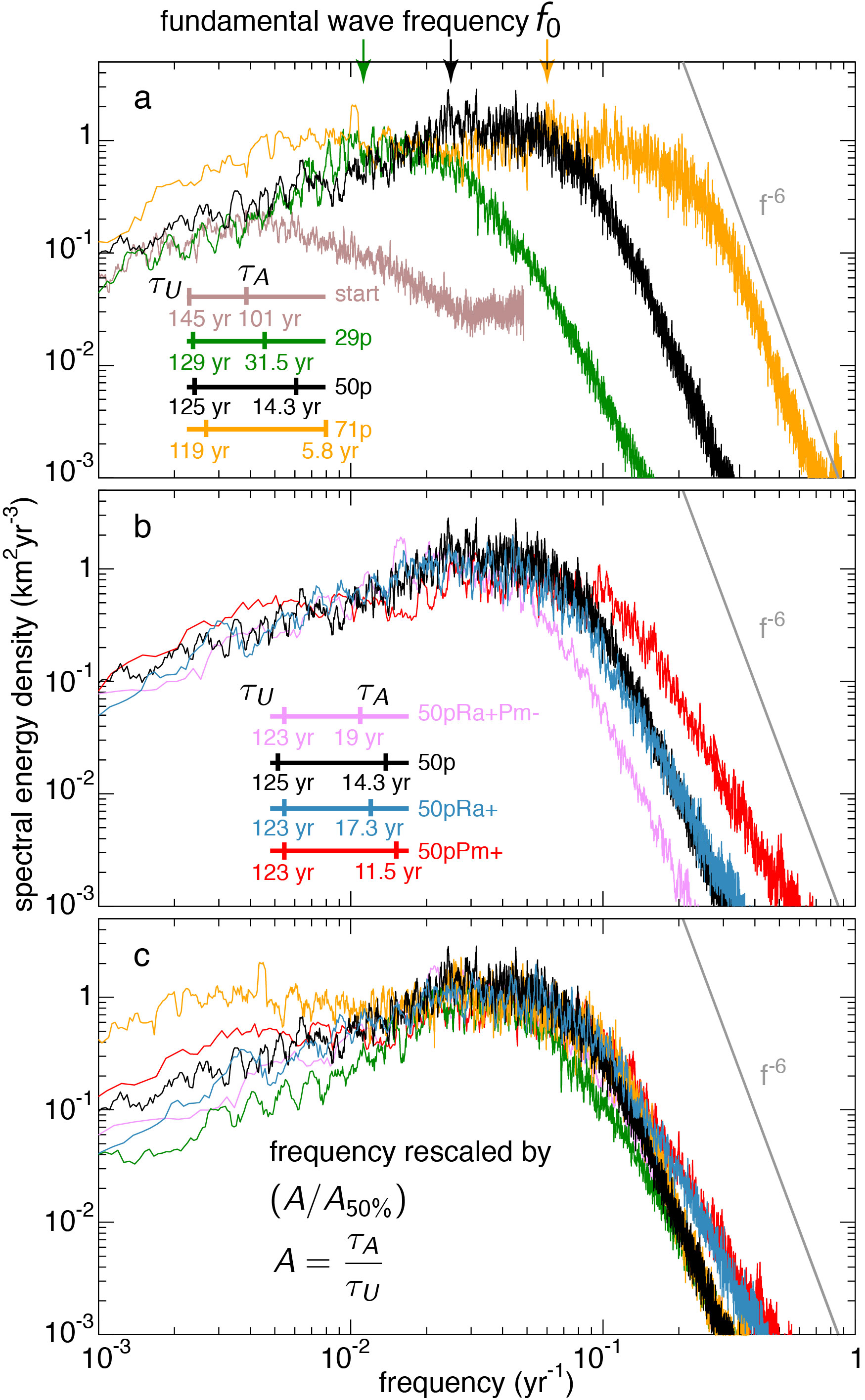}}
\caption{\label{ZA30} Frequency domain spectral density $\hat{K}^\mathrm{ZS}(f)$ of the core surface azimuthal, equator-symmetric and axisymmetric acceleration energy for models taken (a) along and (b) across the parameter space path. a: spectra of the start, 29 percent, 50 percent and 71 percent of path models. b: spectra of the 50 percent of path, 50pRa+Pm-, 50pRa+ and 50pPm+ models. The numerical values of the time scales $\tau_{U}$ and $\tau_{A}$ are also recalled in a,b. In a., the arrows locate the frequency of the fundamental Alfvén wave $f=(\sqrt{3}r_{o}\tau_{A})^{-1}$ for models along the path, and the slanted grey line marks the power law $f^{-6}$. c: compound of a.,b., with frequency axis rescaled for each model by $A/A_{50\%}$, where $A=\tau_{A}/\tau_{U}$ and $A_{50\%}$ is the Alfvén number of the Midpath model.}
\end{figure}

\rev{At the conditions of low inertia and viscosity where our models operate, torsional waves are excited by} perturbations of the Lorentz force averaged over axial cylinders, with these perturbations being linked to the underlying convection \rev{\citep{Teed2014,Teed2019}. This process naturally leads to} excitation of large-scale nature. This explains why almost all of the signal present in $\dot{u}_{g}$ and evaluated from bulk data at native resolution (Fig. \ref{TW71}b) can be retrieved in a time-cylindrical radius diagram of $\dot{u}_{\varphi}^{\mathrm{ZS}}$ (Fig. \ref{TW71}c), the equator-symmetric zonal flow acceleration at the core surface up to spherical harmonic degree 30. In Fig. \ref{ZA30} we further analyse the associated energy density spectra $\hat{K}^\mathrm{ZS}(f)$ in the frequency domain. Beyond 29 percent of the path (Fig. \ref{ZA30}a), most of the signal corresponds to the contribution from torsional waves, with the invariant contribution from axisymmetric thermal winds (estimated by representing the start of path model in Fig. \ref{ZA30}a) being subdominant. In these advanced models, $\hat{K}^\mathrm{ZS}(f)$ features a plateau centered around the fundamental wave frequency $f_{0}=(\sqrt{3}r_{o}\tau_{A})^{-1}$ (the factor $\sqrt{3}$ accounting for the conversion between isotropic and one-dimensional wave velocity), terminated at high frequencies by a $f^{-6}$ decay associated with the limited length scale content of the excitation source, and much steeper than that observed for the full core surface flow in Fig. \ref{psd}b. The evolution of this spectral shape along the path reflects a wave content that remains at large and invariant spatial scales while moving towards higher frequencies as the Alfvén velocity is increased. Using cross-path models (Fig. \ref{ZA30}b) again together with along-path cases, the cut-off frequency is indeed confirmed to extend linearly with the inverse Alfvén number $1/A=\tau_{U}/\tau_{A}$ (Fig. \ref{ZA30}c), and a possible scaling with $S$ can be discarded. The influence of magnetic diffusion is marginal here, because of the large length scales at which the waves are excited, and because radial core surface diffusion is subdominant (recall Fig. \ref{psd}a). Effects of diffusion are only seen in the decay tails in Fig. \ref{ZA30}c, with the less diffusive models 50pPm+ and 50pRa+ showing a slightly less steep decay than the other models. The simple form of the spectra implies that the time averaged energy $[K^\mathrm{ZS}]$ (the integral of the profiles along the frequency axis) scales as
\begin{equation}
[K^\mathrm{ZS}] \sim A^{-1}.\label{TOscaling}
\end{equation}
This means that the waves are able to draw an increasing amount of energy from the underlying convection as the separation between $\tau_{A}$ and $\tau_{U}$ increases. \rev{The leading contribution to this increase of $[K^\mathrm{ZS}]$ comes from the additional frequencies that are made available to wave motion as the fundamental Alfvén frequency $f_{0}$ is increased. Resonant forcing by convection \citep[e.g.][]{Teed2019} does not appear to be the dominant excitation mechanism, because the gap widens along the path between the low and constant frequencies of thermal winds (again represented by the start of path model in Fig. \ref{ZA30}a) and the increasing frequency $f_{0}$. Significant energy levels are however preserved in the intermediate frequency range separating the two, where motion is neither the consequence of convection nor of waves, but rather results from a non-linear interaction between the two, with the dominant non-linearity for energy transfers being the Lorentz force in this case \citep{Teed2014,Aubert2017}. The observation of a broadening flat energy density profile in this interaction range is therefore supportive of a non-linear transfer of energy from convection to waves underlain by the Lorentz stresses.} Though the stabilising control of magnetic dissipation is not present at the axisymmetric scale, as we have shown above, it is expected to indirectly come from the non-axisymmetric contributions to these Lorentz stresses.  

\subsection{Non-axisymmetric waves.}

\begin{figure}
\centerline{\includegraphics[width=14cm]{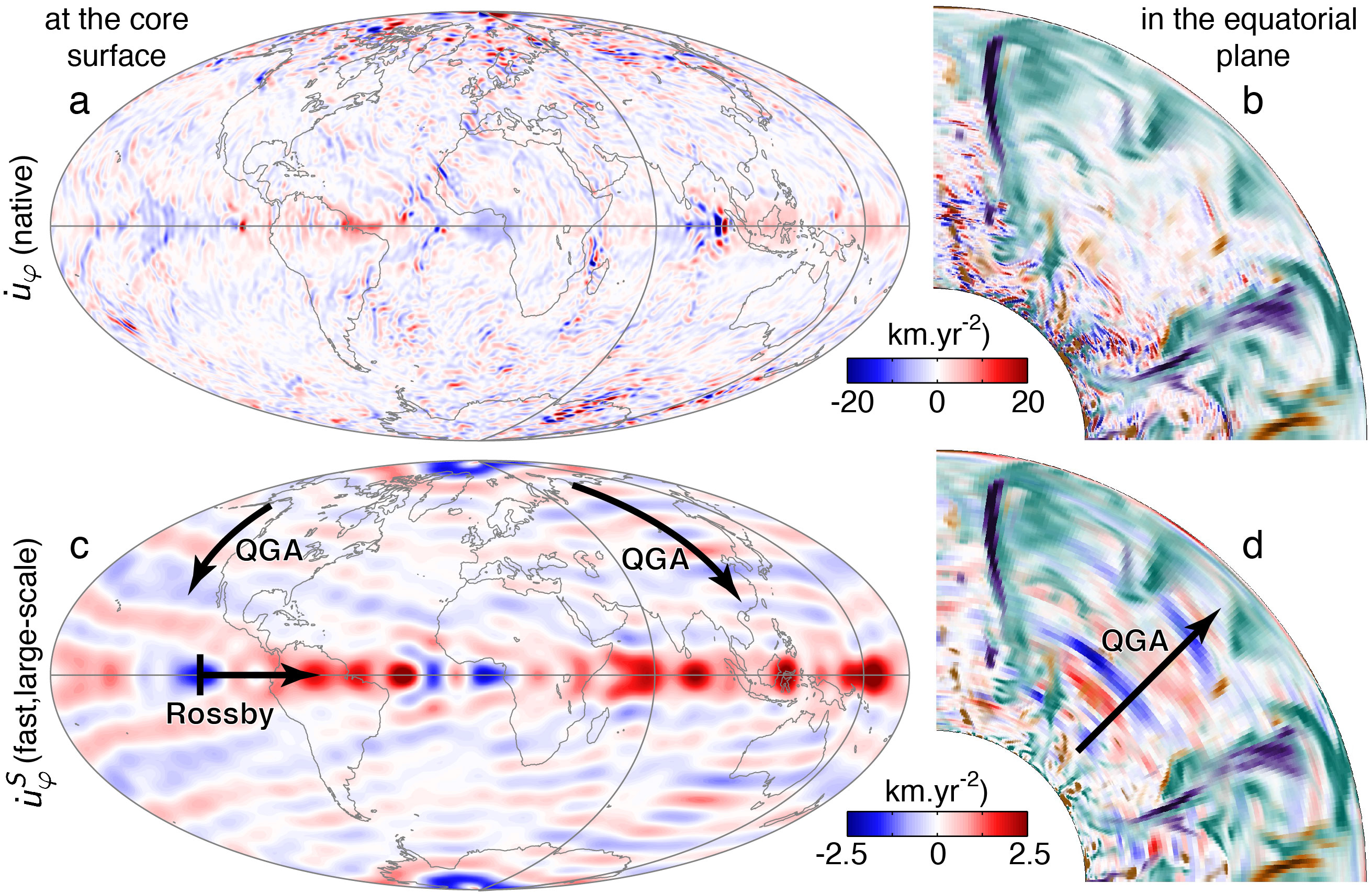}}
\caption{\label{fullfast}State of the 71p model at time 4749.8 yr of the simulation. Core surface Hammer projections (a,c) and partial equatorial plane views (b,d) of the azimuthal flow acceleration. In a,b, the total azimuthal acceleration $\dot{u}_\varphi$ is presented at native spatial and temporal resolution. In c,d, the fast and large-scale component of the equator-symmetric azimuthal acceleration $\dot{u}_\varphi^\mathrm{S}$ is presented. The fast component is obtained after removal of a 5-year running average. The large-scale component is obtained in c. by truncation after spherical harmonic degree 30, and in d. by lateral truncation after azimuthal wavenumber 30 and application of a running average of thickness 40 km in the radial direction. For clarity also, accelerations larger than $4\ut{km.yr^{-2}}$ are not represented in d. The partial equatorial views in b,d, are taken between the grey meridians delineated in a,c. In these equatorial views also, the density anomaly (green-brown, maximal values $\pm 0.5~\te{-4}\ut{kg/m^{3}}$) and radial magnetic field (orange-purple, maximal values $\pm 9.5 \ut{mT}$) are represented in a selectively transparent manner. Black arrows in c,d, denote the propagation direction of quasi-geostrophic Alfvén (QGA) and Rossby waves. The QGA wave arrow in d. also marks the longitude selected for analysis in Fig. \ref{QGA71}.}
\end{figure}

Searching for non-axisymmetric waves is intrinsically more difficult as the signature of convection is stronger than for axisymmetric motion, and can attain short time scales at small spatial scales (see equation \ref{MACts}). The wave signature can therefore not be straightforwardly disentangled from that of convection, \rev{for instance} in native resolution snapshots of the azimuthal flow acceleration $\dot{u}_{\varphi}$ at the core surface (Fig. \ref{fullfast}a) and within the equatorial plane (Fig. \ref{fullfast}b). Focusing on the large-scale component (up to degree 30), where the acceleration is overwhelmingly equator-symmetric i.e. $\dot{u}_{\varphi}\approx \dot{u}_{\varphi}^\mathrm{S}$ (see section \ref{methodseries}) clarifies the picture by partly removing the convective signal originating at small spatial scales. For the purpose of illustration, in Fig. \ref{fullfast}c,d we further consider the rapid part of the large-scale signal, obtained after removal of a 5-year running average. This processing highlights \rev{two other classes} of rapidly propagating features (located by labels in Fig. \ref{fullfast}c,d) in addition to the already described axisymmetric torsional waves\rev{: transverse quasi-geostrophic Alfvén (QGA) waves and longitudinal Rossby waves.}

\subsubsection{Quasi-geostrophic Alfvén waves.}
Non-axisymmetric, \rev{axially columnar} QGA waves propagate \rev{along magnetic field lines of arbitrary orientation perpendicular to the rotation axis. The typical distribution of field lines in the model \citep[see e.g. Fig. 8 in][]{Aubert2019b} mainly promotes cylindrical-radially propagating waves in the bulk of the fluid that are carried by the azimuthal flow $u_{\varphi}$ (similarly to torsional waves). When tracked in the equatorial plane (Fig. \ref{fullfast}d),} these are laterally limited by concentrations of radial magnetic field, as previously described in \cite{Aubert2018}. \rev{Near concentrations of azimuthal magnetic field found at low latitudes, the propagation direction can also acquire an azimuthal component (see e.g. Fig. \ref{fullfast}d near America).} While the QGA waves were previously mostly exhibited in the vicinity of magnetic acceleration pulse events, \rev{they are now significantly stronger in the 71p model (see section \ref{wavenerg}) and we can} confirm that they are ubiquitous and constantly sent out by the deep convection. 

\begin{figure}
\centerline{\includegraphics[width=14cm]{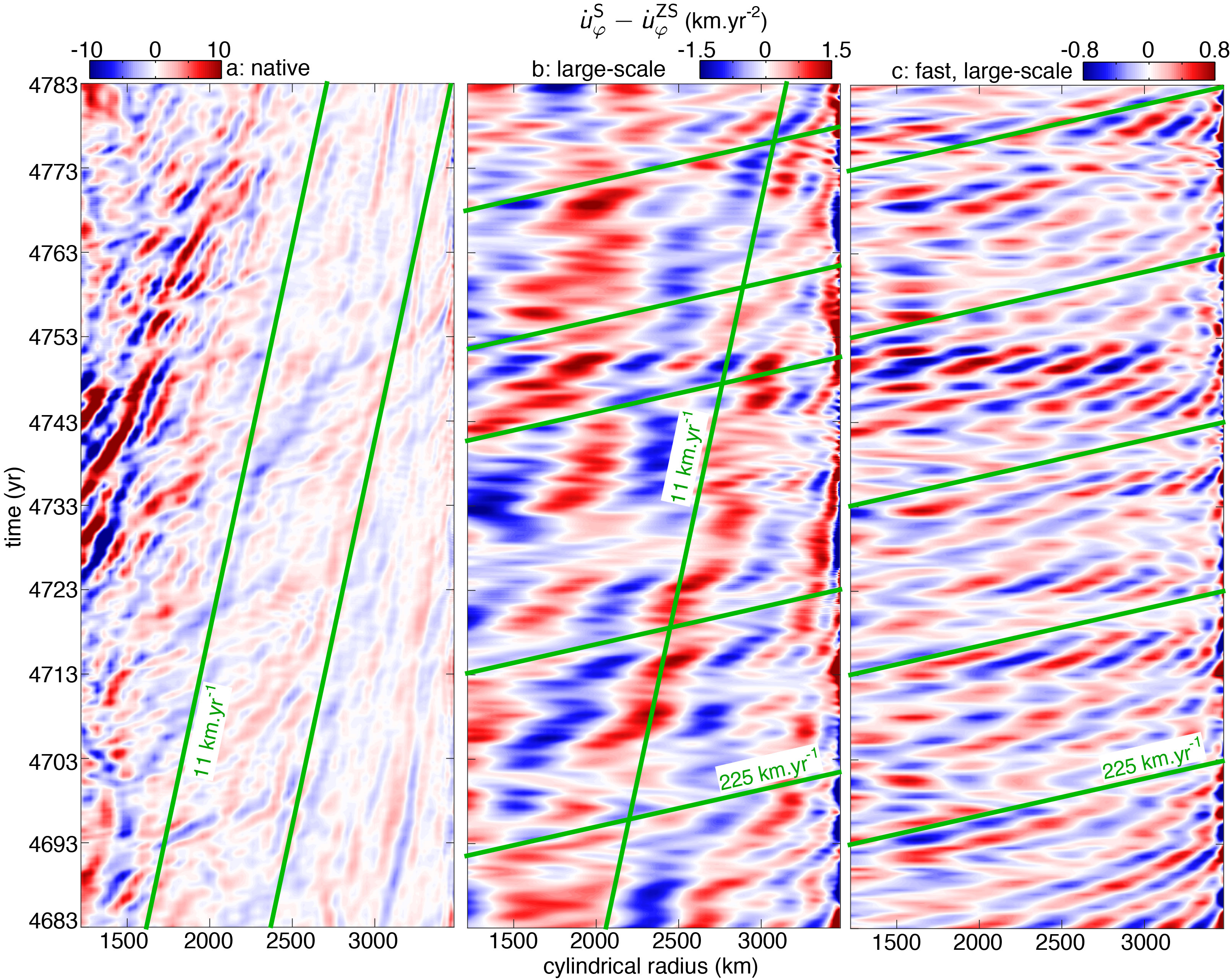}}
\caption{\label{QGA71}Time-cylindrical radius diagrams of the non-axisymmetric, equator-symmetric core surface azimuthal flow acceleration $\dot{u}^\mathrm{S}_\varphi-\dot{u}^\mathrm{ZS}_\varphi$ at longitude $112.5^\mathrm{o}\mathrm{E}$ (located by arrows in Fig. \ref{fullfast}c,d) in the 71p model, presented at (a) native spatial and temporal resolution, (b) large-scale (spherical harmonic degree up to 30), and (c) large-scale for the fast component obtained after removal of a 5-year running average in time. The sets of slanted green lines denote advection at the one-dimensional convective velocity $D/\sqrt{3}\tau_{U}= 11 \ut{km/yr}$, and propagation at the one-dimensional Alfvén wave velocity $D/\sqrt{3}\tau_{A}= 225 \ut{km/yr}$.}
\end{figure}

To better characterise the propagation of QGA wave features at the core surface in the 71p model, in Fig. \ref{QGA71} we now specifically focus on the non-axisymmetric, azimuthal and equator-symmetric flow acceleration $\dot{u}^\mathrm{S}_\varphi-\dot{u}^\mathrm{ZS}_\varphi$ and examine time-cylindrical radius diagrams at a specific longitude delineated by an arrow in Fig. \ref{fullfast}c,d. Representing this quantity at native spatial and temporal resolution (Fig. \ref{QGA71}a) mainly reveals the outwards advection of flow by convection at speeds commensurate with the one-dimensional convective velocity  $D/\sqrt{3}\tau_{U}= 11 \ut{km/yr}$. On this representation, however, faint signals propagating outwards at higher speeds can already be seen. In the large-scale component (Fig. \ref{QGA71}b) that removes the signature of fast convective signatures, we now observe together the slow convective and faster wave signals. Singling out these latter signals by again removing a 5-year running average in time (Fig. \ref{QGA71}c) reveals QGA waves with outward propagation speeds close to, but lower than the one-dimensional Alfvén velocity $D/\sqrt{3}\tau_{A}= 225 \ut{km/yr}$. Comparing Figs. \ref{QGA71}c and \ref{TW71}c also shows that QGA waves can reach significantly smaller time scales and hence shorter radial wavelengths than torsional waves. 

While the Coriolis force identically vanishes from the magneto-inertial balance that drives axisymmetric torsional waves, it is necessarily present at the non-axisymmetric level and fundamentally modifies the wave equation \cite[e.g.][]{Finlay2008houches,Finlay2010}. This leads in particular to magneto-Coriolis waves with phase velocities much smaller than the Alfvén velocity. These are furthermore dispersive, with the wave velocity increasing as the wavelength decreases. The effects of the Coriolis force can however be mitigated for axially-invariant waves of sufficiently small wavelength, that propagate in a direction perpendicular to the rotation vector. In this case, the wave velocity can approach the Alfvén velocity from below \citep{Gerick2020}, as the influence of inertia is restored within the magneto-Coriolis balance. The axially columnar structure, propagation direction perpendicular to the rotation vector, velocities lower than $D/\sqrt{3}\tau_{A}$ and short radial wavelengths of the QGA waves observed here can also be understood within this framework. \rev{Our configuration however differs from that of \cite{Gerick2020} in the sense that the background magnetic field is highly complex, with relatively homogeneous regions surrounded by strongly heterogeneous field line concentrations corresponding to sharp gradients \citep[e.g. Fig. \ref{fullfast}b,d, see also Fig. 8 in][]{Aubert2019b}. This situation promotes a spatial segregation of the force balance \citep{Aubert2018}. The magneto-Coriolis part of this balance is mostly observed at the edges of the QGA wavefronts which correspond to the locations of strongest magnetic field heterogeneity. In the more magnetically homogeneous regions, however, the Coriolis force is mitigated to the point where wave motion becomes locally magneto-inertial again.} The residual \rev{influence} of the Coriolis force \rev{nevertheless tends to promote}  smaller radial length scales for QGA waves relative to torsional waves, which also rationalise the different behaviours of energy density spectra $\hat{K}(f)$ and $\hat{K}^\mathrm{ZS}(f)$ (compare Figs. \ref{psd}b,d,f and \ref{ZA30}) as regards the steepness of their high-frequency decay and the influence of bulk magnetic diffusion (or lack thereof) on this decay. \rev{We also expect this residual influence to increase as the waves approach the core surface at equatorial position, because of the increasing slope of the spherical boundary. This leads to additional wave slowdown compared to torsional waves and hence (the wave period being preserved) further reduction of the radial length scale (compare again Figs.  \ref{QGA71}c and \ref{TW71}c.)} 

\subsubsection{\label{RW}Rossby waves}
\rev{The Rossby waves observed in Fig. \ref{fullfast}c propagate eastwards in the vicinity of the equatorial plane. These are slow inertial waves in the sense} that their pulsation $\omega$ is much smaller than the rotation rate $\Omega$, thereby ensuring an evolution under a quasi-geostrophic force balance and an axially columnar structure \citep{Zhang2001,Busse2005,Canet2014}. \rev{From a geomagnetic standpoint, they are however fast in the sense that $\omega$ is larger than the typical pulsation $2\pi/\tau_{A}$ of Alfvén waves. As a result, they are} only slightly modified by the presence of the magnetic field \citep[e.g.][]{Finlay2010} and typically feature magnetic to kinetic energy ratios smaller than unity \cite[e.g.][]{Gerick2020}. \rev{They hence do not} bear a significant signature in the magnetic acceleration signals of Fig. \ref{psd}, where a possible control from $\tau_{\Omega}$ is elusive. 

\begin{figure}
\centerline{\includegraphics[width=14cm]{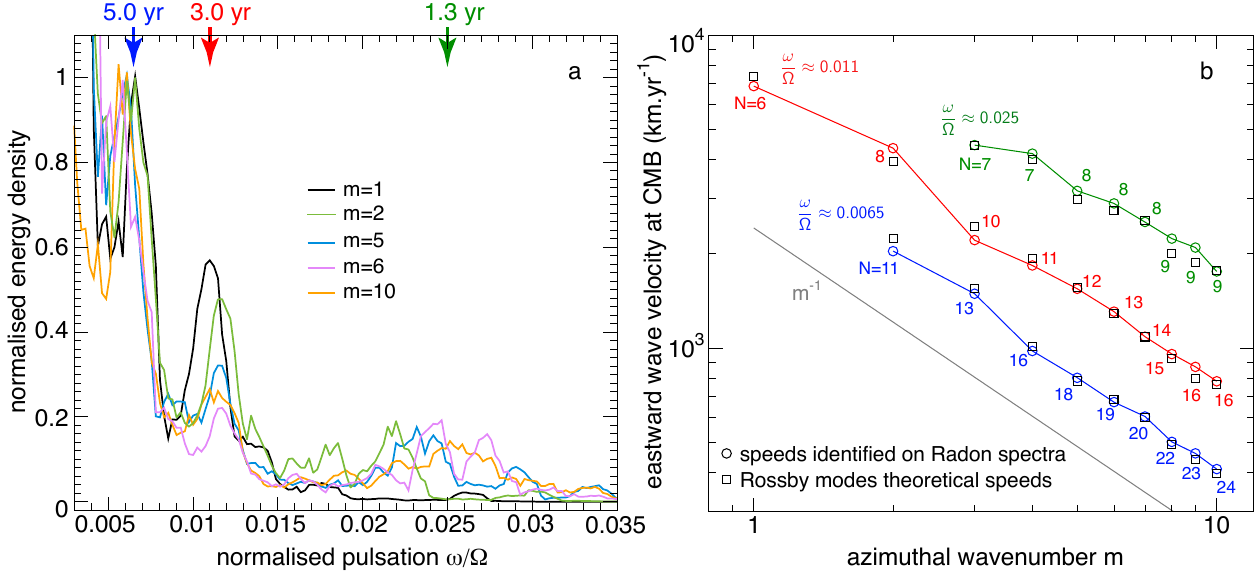}}
\caption{\label{Rossby}a: Frequency-domain energy density profiles of $\dot{u}_\varphi$ at the equator in a 130-year sequence of the 71p model. The acceleration signal is broken down into contributions from distinct azimuthal mode numbers $m$ and the energy is normalised relatively to the peak at $\omega/\Omega\approx0.0065$. Also reported at the top of the graph are the periods corresponding to each of the three identified peaks (located by colored arrows). b: Eastward propagation velocities (circles) identified on Radon-transformed time-longitude diagrams of equatorial azimuthal flow acceleration \citep[following the procedure described in][]{Finlay2005}, represented as a function of the mode number $m$ at each of the three normalised pulsations $\omega/\Omega\approx 0.0065, 0.011, 0.025$ corresponding to the peaks in panel a (see arrows of same color). Also reported as functions of $m$ are the closest matching phase velocities $c=\omega r_{o}/m$ (squares) obtained by adjusting the radial complexity number $N$ in equation (\ref{oO}). The grey line represents a $m^{-1}$ law.}
\end{figure}

In a full sphere \citep{Zhang2001} \rev{(a configuration that reasonably applies to our situation where the Rossby waves are confined at low to mid-latitudes, Fig. \ref{fullfast}c)}, and when $\omega/\Omega<<1$, the normalised pulsation $\omega/\Omega$ of \rev{hydrodynamic} Rossby waves can be approximated by the expression 
\begin{equation}
\dfrac{\omega}{\Omega}\approx\dfrac{2}{m+2}\left[\sqrt{1+\dfrac{m(m+2)}{N(2N+2m+1)}}-1\right],\label{oO}
\end{equation}
which involves the azimuthal mode number $m$ as well as a number $N$ describing their level of complexity in the radial direction. Fig. \ref{Rossby}a presents frequency-domain energy density profiles of the equatorial acceleration seen in Fig. \ref{fullfast}a, broken down into contributions from distinct azimuthal mode numbers $m$. The peaks that can be observed at various mode numbers are nearly synchronised in the vicinity of several distinct normalised pulsations. We focus on the three normalised pulsations $\omega/\Omega \approx 0.0065, 0.011, 0.025$, which contain significant energy across different values of $m$. The eastward propagation velocity of the corresponding waves can be determined by performing a Radon transform on time-longitude diagrams of the fast equatorial acceleration signal, again broken down by mode number $m$. At each value of $m$, Fig. \ref{Rossby}b reports the propagation velocities identified at distinct Radon energy peaks. Next to observed velocities, Fig. \ref{Rossby}b also reports theoretical phase velocities $c=\omega/k$, where $k=m/r_{o}$ and $\omega$ is determined from equation (\ref{oO}) by finding the radial complexity level $N$ that best matches the observed velocity. The agreement between observed and theoretical velocities is excellent throughout the range investigated for $\omega/\Omega$ and $m$, and provides an unambiguous characterisation of the Rossby waves seen in Fig. \ref{fullfast}a (and more clearly in Fig. \ref{fullfast}c after filtering out the low-frequency content). The synchronisation of $\omega/\Omega$ across distinct mode numbers $m$ is possible because despite their discrete character, the possible values given by equation (\ref{oO}) form a ensemble that is dense enough to offer $(m,N)$ couples yielding a pulsation close to $\omega/\Omega \approx 0.0065, 0.011, 0.025$. The synchronisation is then achieved by non-linear crosstalk between the waves, which is also necessary to ensure a saturation of their amplitude. Because of this synchronisation leading to nearly constant $\omega/\Omega$, the phase speed $c=\omega r_{o}/m$ of each wave category in Fig. \ref{Rossby}b naturally scales like $m^{-1}$. Unlike Alfvén waves, Rossby waves are indeed highly dispersive in nature, with small-scale waves being significantly slower than large-scale waves. While all the waves observed here have periods within the interannual time scale range (see arrows in Fig. \ref{Rossby}a), we note that at the largest scales, their velocities up to $10^{4} \ut{km/yr}$ are the fastest observed in the 71p model. 

\subsection{\label{wavenerg}Wave energy scaling along the path.}
\begin{figure}
\centerline{\includegraphics[width=9cm]{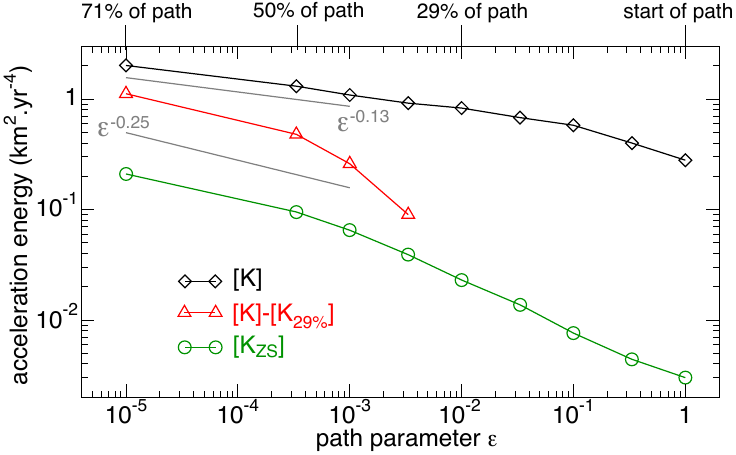}}
\caption{\label{ve}a: Evolution with the path parameter $\epsilon$ of the time-averaged, total core surface flow acceleration energy $[K]$ up to spherical harmonic degree 30, plotted together with two diagnostics of the evolution of wave energy along the path (Earth's core conditions are towards the left of the graph). The energy of axisymmetric torsional waves is represented by the axisymmetric, azimuthal and equator-symmetric part $[K^\mathrm{ZS}]$, and the energy of non-axisymmetric waves is represented by the difference $[K]-[K_{29\%}]$ of $[K]$ to its value at 29 percent of the path. The grey lines mark the $\epsilon^{0.13}$ and $\epsilon^{0.25}$ power-law behaviours.}
\end{figure}

In Fig. \ref{ve} we present the evolution of flow acceleration energy along the parameter space path for the various waves that we have isolated. As measured by $[K^\mathrm{ZS}]$, the energy of torsional waves remains subdominant relatively to the total acceleration energy $[K]$, but gains two orders of magnitude along the path to reach $[K_{ZS}]/[K] \approx 0.1$ at 71 percent. At advanced path positions, the scaling $[K^\mathrm{ZS}] \sim A^{-1} \sim \epsilon^{-1/4}$ from equation (\ref{TOscaling}) is well approached by our results. in contrast,  $[K]$ asymptotically evolves less steeply (with $[K]\sim \epsilon^{-0.13}$ from our results) because it contains an invariant contribution from convection. This contribution is approximately removed by subtracting to $[K]$ the acceleration energy $[K_{29\%}]$ at the entry into the asymptotic regime, and $[K]-[K_{29\%}]$ may be seen as the total wave contribution that includes non-axisymmetric motion. The spectral form (\ref{Kscal}) suggests the scaling $[K]-[K_{29\%}] \sim S \rev{\sim A^{-1}Rm} \sim\epsilon^{-1/4}$, again in \rev{fair}  agreement \rev{(albeit being clearly validated by two points only)} with our numerical data at advanced path positions. At 71 percent of the path, the total wave contribution $[K]-[K_{29\%}]$ amounts to about half the total flow acceleration energy $[K]$, and is five times stronger than the energy $[K^\mathrm{ZS}]$ of axisymmetric torsional waves. If we denote as $\dot{U}_{W}$ the typical acceleration of waves in the total acceleration $\dot{U}$, in our advanced models this means that 
\begin{equation}
\dot{U}_{W}/\dot{U} \approx \left(\dfrac{[K]-[K_{29\%}]}{[K]}\right)^{1/2}=\mathcal{O} (1).\label{UdotWUdot}
\end{equation}

We terminate our analysis by estimating the contributions of waves to the core flow and magnetic field (and no longer to the time derivatives of these quantities, as we have done so far). Because these are relatively small, as we shall see, and because of the difficulties in disentangling the waves from convection that we have already alluded to, this cannot be straightforwardly achieved from the numerical model output and we need to take an alternative route. Assuming that the waves mainly evolve with time scale $\tau_{A}$, while the convection evolves with time scale $\tau_{U}$ (with $A=\tau_{A}/\tau_{U}$), and denoting again as $U_{W}$ the contribution of waves to the typical velocity field amplitude $U$, we estimate that
\begin{equation}
U_{W}/U=A \dot{U}_{W}/\dot{U}= \mathcal{O} (A).\label{UWU}
\end{equation}
We can further assume that the typical magnetic field amplitude $B_{W}$ carried by the Alfvén waves obeys energy equipartition i.e. $B_{W}/U_{W}= \sqrt{\rho\mu}$. Using the definition $A=\sqrt{\rho\mu}U/B$, we then also estimate that
\begin{equation}
B_{W}/B=\dfrac{B_{W}}{U_{W}}\dfrac{U_{W}}{U}\dfrac{U}{B}=\mathcal{O}(A^{2})\label{BWB}
\end{equation}
This estimate is in principle valid at any altitude above the core-mantle boundary, if we assume that waves and convection share a similar length scale content. Equations (\ref{UWU},\ref{BWB}) show that the wave signatures in the total flow and magnetic fields are very subdominant since $A\ll 1$, and confirm that differentiating the velocity field once in time and the magnetic field twice in time is essential to highlighting these contributions. 

\section{\label{discu}Discussion}
\subsection{Waves and convection in the frequency domain}
In numerical models sampling the parameter space path introduced in \cite{Aubert2017}, dynamics of convective and wave origin occupy different, but overlapping frequency ranges, as seen for instance in spectral energy density profiles of the magnetic and flow acceleration (Fig. \ref{psd}). Of particular interest is the secular to decadal range located between the (approximately constant) overturn frequency $f_{U}=1/\tau_{U}$ and the cut-off frequency $f_{c}$, where energy density presents a plateau. This is where slow convection and rapid waves are in potential interplay. We have seen that slow convection accounts for the leftmost part of this plateau, the invariance of which indeed mirrors the kinematic invariance observed along the path. It is possible that this plateau relates with vorticity equivalence in the spatial domain \citep{Davidson2013,Aubert2019b}, which states that vorticity tends to be evenly distributed by magnetic turbulence in the range $[d_\mathrm{min},d_{\perp}]$ between the scale of magnetic dissipation and the dominant length scale at which convection powers the magnetic field. Assuming a simple linear correspondance between frequencies and length scales, this would indeed imply $u(f) \sim f^{-1}$ and $\dot{u}(f) \sim f^{0}$. This reasoning is appealing for explaining the leftmost end of this plateau but falls short of interpreting the rightmost end, the flattening of which along the path is a consequence of an elevating contribution coming primarily from hydromagnetic Alfvén waves. This was confirmed in particular by the control on the cut-off frequency $f_{c}$ of the Lundquist number $S$ (equation \ref{cutoff}), which measures the number of Alfvén wave periods occurring over the course of a magnetic diffusion time. The case of torsional waves (Fig. \ref{ZA30}) helped to illustrate the spectral broadening that can be expected as the time scale separation between the waves and convection increases, because of the higher wave frequencies that are made available and the non-linear wave-convection interactions underlain by Lorentz stresses. We hence conclude that the flat magnetic acceleration energy density profile in the range $[f_{U},f_{c}]$ \citep[corresponding to the $f^{-4}$ range observed in energy density spectra of the geomagnetic field,][]{DeSantis2003,Lesur2018} can be ascribed to the combined effect of non-linearly interacting slow convection and rapid Alfvén waves. 

At higher frequencies $f \ge f_{c}$, most of the wave range is characterised by a $f^{-2}$ decay power law. In this range, the wave acceleration energy is controlled by the Lundquist number $S=\tau_{\eta}/\tau_{A}$ (equations \ref{Mscal},\ref{Kscal}), again highlighting the importance of time scale separation. This also underlines the stabilising control of magnetic diffusion felt by waves in the bulk of the fluid (and not at the core surface, Fig. \ref{psd}). Diffusion is essentially active at non-axisymmetric scales, where waves typically reach small radial length scales mitigating the influence of the Coriolis acceleration. In terms of length scales also, the dependence $f_{c}\sim \sqrt{S}$ (equation \ref{cutoff}) highlights the importance of the bulk magnetic diffusion length scale at the Alfvén time scale $\delta=\sqrt{\eta\tau_{A}}=D/\sqrt{S}$. Waves at the largest scales such as axisymmetric torsional waves are immune to diffusion (Fig. \ref{ZA30}) but are nevertheless limited in amplitude because they are powered by non-linear couplings that involve non-axisymmetric scales where diffusion-limited Alfvén waves are found. As we advance along the parameter space path, and despite the tightening constraints set by the QG-MAC balance, the availability of higher wave frequencies and the non-linear energy transfers cause an increase in the total wave acceleration energy (Fig. \ref{ve}). At 71 percent of the path we evaluate the total contribution from waves to the flow acceleration energy to be equivalent to that of convection when considering the signal up to spherical harmonic degree 30. At a fifth of the wave acceleration energy, the contribution from torsional waves is subdominant.

Using values $\tau_{\eta}=135000 \ut{yr}$, $\tau_U=130\ut{yr}$ and $\tau_{A}=2\ut{yr}$ (table \ref{outputs1}b), at Earth's core conditions we predict $S \approx 7~\te{4}$ and \rev{$f_{c} \approx 0.07 \ut{yr^{-1}}$} from the scaling (\ref{cutoff}). \rev{The $f^{0}$ range of interplay between waves and convection should therefore extend over a decade only between $f_{U}=8~\te{-3} \ut{yr^{-1}}$ and $f_{c}$. The Alfvén frequency $1/\tau_{A}=0.5\ut{yr^{-1}}$ is predicted to lie well into the $f^{-2}$ range, at an acceleration energy density level of about $(f_{c}\tau_{A})^{2}\approx 1/50$ below the plateau (from equations \ref{Mscal},\ref{Kscal}). Assuming $Rm\approx 2000$ in the core (by doubling $\tau_{\eta}$) rather than the value $Rm\approx 1000$ adopted along the path shifts the cut-off frequency to $f_{c}\approx 0.1 \ut{yr^{-1}}$ and the relative attenuation at the Alfvén frequency to $1/25$. These last results fall} short of accounting for the observed extent of the $f^{0}$ acceleration ($f^{-4}$ magnetic) range up to frequency $f \approx 1\ut{yr^{-1}}$ \citep{Lesur2018}, \rev{such that more fundamental modifications of the modelling set-up appear to be needed.} The path models are possibly underpowered at high frequencies because their design involves a neutrally buoyant core-mantle boundary. Exploring the effects of convective instability near the core surface on the high-frequency content of magnetic acceleration may therefore provide important geophysical constraints. It is unlikely that any degree of upper outer core stratification would help to bring the numerical results close to the geomagnetic observation, as this would further hinder the short-timescale dynamics \citep{Aubert2019}. The stratification would also degrade the morphological resemblance of the model core surface magnetic field to the present-day field \citep{Gastine2020}. From the observational standpoint, this discussion also highlights the importance of efforts aiming at cleaning the geomagnetic acceleration signal from external contributions close to annual frequencies \citep[see e.g.][]{Finlay2017}. 

\subsection{Detectability of hydromagnetic waves in geomagnetic observations.}
Equation (\ref{BWB}) suggests that the relative contribution $B_{W}$ from waves in the magnetic field amplitude $B$ is $B_{W}/B=\mathcal{O}(A^{2})$ i.e. $B_{W}\approx 8\ut{nT}$ at Earth's surface and at the physical condition $A=1.5~\te{-2}$ of the core. Though this may appear extremely weak, it remains within the typical resolution of satellite-based observations \citep[see e.g.][]{Finlay2016b}, and we have seen (Figs. \ref{psd},\ref{ve}, equation \ref{UdotWUdot}) that differentiating the magnetic field twice in time, or the velocity field once in time, leads to sizeable wave contributions to the total acceleration energy. This underlines the crucial importance of extracting the geomagnetic acceleration from ground observatory and satellite data at a good level of spatial and temporal resolution. This also motivates research towards elaborate approaches aiming at extracting a reliable flow acceleration from the magnetic acceleration signal \citep[see a recent review in][]{Gillet2019b}. Furthermore, the increase of wave energy along the path (Fig. \ref{ve}), the ubiquitous character of waves (Figs. \ref{fullfast},\ref{QGA71}) as well as the large-scale content of torsional waves (Fig. \ref{TW71}) appear to promote some optimism as regards the detectability of waves in the geomagnetic signal emanating from Earth's core.

Among the signal caused by waves in the magnetic acceleration, hydromagnetic Alfvén waves are dominant because (unlike high-frequency Rossby waves) they achieve equipartition between the kinetic and magnetic energies that they carry (Fig. \ref{psd}). These should hence be most straightforwardly detectable in the geomagnetic acceleration signal. Concerning the typical flow amplitude $U^\mathrm{ZS}$ of torsional waves, following the estimation strategy leading to equation (\ref{UWU}) we infer $U^{ZS}/U \approx A \sqrt{[K^\mathrm{ZS}]/[K]}$, with an extrapolation to the end of the path yielding $[K^\mathrm{ZS}]/[K] \approx 0.17$. Using $U=D/\tau_{U} \approx 17 \ut{km/yr}$ and $A=1.5 ~\te{-2}$, this finally leads to $U^\mathrm{ZS} \approx 0.1 \ut{km/yr}$, somewhat smaller than (but of the same order of magnitude as) the value $\approx 0.3 \ut{km/yr}$ obtained from modelling of the rapidly evolving core flow \citep[see Fig. 13 in][]{Gillet2015}. Likewise, at the end of the path the total wave amplitude $U_{W}$ including axisymmetric and non-axisymmetric contributions should be higher. From equations (\ref{UdotWUdot},\ref{UWU}) we infer $U_{W}/U \approx A \sqrt{([K]-[K_\mathrm{29\%}])/[K]}\approx A$ and therefore $U_{W} \approx 0.25 \ut{km/yr}$, again comparable to the value $\approx 0.6 \ut{km/yr}$ obtained by \cite{Gillet2015} for the non-zonal flow. We finally note that the value $[K^\mathrm{ZS}]/([K]-[K^\mathrm{ZS}]) \approx 0.2$ that we predict at the end of the path for the ratio of zonal to non-zonal power in the core flows is in fair agreement with that retrieved by \cite{Gillet2015} at interannual frequencies (their Fig. 9). Our best chances to characterise the signature of non-axisymmetric Alfvén waves still rests in the analysis of the geomagnetic jerks that they cause, because of the associated geometrical wave amplification effects \citep{Aubert2019}. We have found stronger and more frequent jerks in the 71p model than at earlier path positions, though some jerks events may become too fast for being noticeable in determinations of the geomagnetic acceleration with limited temporal resolution (Figs. \ref{SVSAmorph},\ref{EJ}). Previously, we had ascribed the increase of jerk energy $E_\mathrm{J}$ along the path to an increased level of wave radial shoaling, but the present results rather incite us to simply associate it with the linear increase of wave acceleration energy with $S$ that we have documented in this study.

The case for detectability of hydrodynamic Rossby waves in the geomagnetic signal appears considerably less obvious, as we did not find a conclusive influence of the rotational time scale $\tau_{\Omega}$ in the magnetic acceleration energy spectra of our models (Fig. \ref{psd}). Only the slowest Rossby modes with periods approaching the Alfvén time scale $\tau_{A}$ can in principle be detected from geomagnetic observations, because faster modes carry less magnetic than kinetic energy \citep[e.g.][]{Gerick2020}. Although these modes also feature the slowest eastward propagation speeds (Fig. \ref{Rossby}), these speeds are still up to several thousands of kilometers per year in the 71p model and should even be ten times faster in the core as they scale linearly with $1/\tau_{\Omega}$ (equation \ref{oO}). Seen from the standpoint of typical geomagnetic acceleration timescales, these would therefore amount to almost instantaneous signals and would hence be very hard to isolate as propagating features, unless the signal is considered at the smallest spatial scales where the velocity of these dispersive waves is considerably slower. \rev{Eastward-propagating, equatorial} geomagnetic signals at speeds \rev{in the range} $500-1000\ut{km/yr}$ \rev{and wavenumbers $m=3-7$} have recently been inferred from geomagnetic acceleration records \citep{Chulliat2015,ChiDuran2020}. \rev{According to our analysis, at Earth's core conditions these could be explained in terms of Rossby waves only if the spatial complexity level $N$ (equation \ref{oO}) was allowed to reach high values $N> 60$, which is beyond the latitudinal resolution available for geomagnetic acceleration.}

\subsection{The path theory in the light of the 71p model and future prospects towards reaching Earth's core conditions}
Following the path theory introduced in \cite{Aubert2017}, this study has introduced a model located at 71 percent of this path. This has considerably enlarged the asymptotic portion (beyond 29 percent) where similar models are available, and the 71p model could again provide a complete validation to this theory. The leading order QG and first-order MAC force balances are stable, while the amplitude of inertia and viscosity continually decrease along the path \rev{\citep[compare Fig. \ref{forcebal} to Fig. 1 of][]{Aubert2018}}. Power-driven, diffusivity-free scaling laws proposed in \cite{Aubert2017} could be confirmed (Fig. \ref{scalings}). In the 71p model this leads to a state where an overwhelming fraction of the injected convective power is Ohmically dissipated, and where the Taylor constraint is enforced at a high level (Table \ref{outputs2}). The dynamo is also in a strong-field state, with the ratio of magnetic to kinetic energy being directly given by $1/A^{2}=416$, now just an order of magnitude away from Earth's core value. The statics (main field morphology), kinematics (convective core flows, magnetic secular variation), as well as the dynamics at time scales shorter than the convective overturn time $\tau_{U}$ are still invariant and carried over from the start of path, while the short-timescale dynamics is gradually enriched in a way that has been documented here in detail (Fig. \ref{SVSAmorph},\ref{psd}). Here we have demonstrated that most, and therefore probably all of the parameter space path is devoid of abrupt physical transitions. This further rationalises the relevance of earlier dynamo models located close to the start of the path to describe the geodynamo, and this also further strengthens the likeliness of core dynamics being in a similar dynamical regime as that observed at 71 percent of the path. 

With a computation carried out over the course of two years, involving 246 million time steps and 15 million single-core CPU hours, the model at 71 percent of the path may be seen as an extreme endeavour that apparently obscures the prospect of being able to terminate the exploration of this path. A positive note is that the computation of this model provided the opportunity to perform several low-level optimisations in the numerical code towards a faster execution at given number of cores and also towards the possibility to use more cores while maintaining a good strong parallel scaling. Together with an improved generation of supercomputers provided by GENCI in France, we could achieve at least a fivefold increase in computation speed for the path models. This was used to increase the spatial resolution and decrease the hyperdiffusivity at which the 71p model runs, and also 
opened the way to an exploration of cross-path models at 50 percent of the path in a way that could not have been feasible only two years ago, at the time of our earlier study \citep{Aubert2018}. We have also seen that it is no longer necessary to advance in half-decades of $\epsilon$ along the path, and that much larger leaps can be achieved through a simple re-scaling procedure of checkpoint files. The next stop along the road is therefore most probably the final one, i.e. being able to simulate the geodynamo exactly in Earth's core conditions. This will presumably imply another round of optimisations and another generation of Tier-1 supercomputers, but we can in principle foresee the completion of this challenge in the coming decade.

\section*{Acknowledgements}
The authors thank two anonymous referees for comments and Dominique Jault, Nathanaël Schaeffer and Thomas Gastine for discussions and help in code optimisation. JA acknowledges support from the Fondation Simone et Cino Del Duca of Institut de France (2017 Research Grant). This project has also been funded by ESA in the framework of EO Science for Society, through contract 4000127193/19/NL/IA (SWARM + 4D Deep Earth: Core). NG was partially supported by the French Centre National d’Etudes Spatiales (CNES) for the study of Earth’s core dynamics in the context of the Swarm mission of ESA. Numerical computations were performed at S-CAPAD, IPGP and using HPC resources from GENCI-TGCC and GENCI-CINES (Grant numbers A0060402122 and A0080402122).

\section*{Data availability}
The numerical code and simulation data analysed in this study are available from the corresponding author upon reasonable request. The core surface data from the 71 percent of path model is also available online at the URL \\
https://4d-earth-swarm.univ-grenoble-alpes.fr/data. 

\bibliographystyle{gji}
\bibliography{Biblio}

\end{document}